\documentclass[a4paper,fleqn,usenatbib]{mnras}

\usepackage[T1]{fontenc}
\usepackage{ae,aecompl}
\usepackage{graphicx}
\usepackage{amsmath} 
\usepackage{amssymb}

\usepackage{bm}
\usepackage{comment}
\usepackage{color}
\usepackage[caption=false]{subfig}

\specialcomment{note}{\begingroup\color{red}}{\endgroup}

\definecolor{ForestGreen}{rgb}{0.3,0.7,0.3}

\newcommand{\x}{\mathbf{x}}

\includecomment{comment}

\title[Dark-matter clustering with Gaussianization]{Recovering dark-matter clustering from galaxies with Gaussianization}
\author[N. McCullagh et al.]{
Nuala McCullagh,$^{1}$
Mark Neyrinck,$^{2}$
Peder Norberg$^{1,3}$
and Shaun Cole$^{1}$
\\
$^{1}$Institute for Computational Cosmology, Department of Physics, Durham University, South Road, Durham DH1 3LE, UK\\
$^{2}$Henry A. Rowland Department of Physics and Astronomy, The Johns Hopkins University 3400 N Charles St., Baltimore, MD 21218, USA\\
$^{3}$Centre for Extragalactic Astronomy, Department of Physics, Durham University, South Road, Durham DH1 3LE, UK
}

\begin{document}
\label{firstpage}
\pagerange{\pageref{firstpage}--\pageref{lastpage}}
\maketitle

\begin{abstract}
The Gaussianization transform has been proposed as a method to remove the issues of scale-dependent galaxy bias and nonlinearity from galaxy clustering statistics, but these benefits have yet to be thoroughly tested for realistic galaxy samples. 
In this paper, we test the effectiveness of the Gaussianization transform for different galaxy types by applying it to realistic simulated blue and red galaxy samples. We show that in real space, the shapes of the Gaussianized power spectra of both red and blue galaxies agree with that of the underlying dark matter, with the initial power spectrum, and with each other to smaller scales than do the statistics of the usual (untransformed) density field. However, we find that the agreement in the Gaussianized statistics breaks down in redshift space. We attribute this to the fact that red and blue galaxies exhibit very different fingers of god in redshift space. After applying a finger-of-god compression, the agreement on small scales between the Gaussianized power spectra is restored. We also compare the Gaussianization transform to the clipped galaxy density field and find that while both methods are effective in real space, they have more complicated behaviour in redshift space. Overall, we find that Gaussianization can be useful in recovering the shape of the underlying dark matter power spectrum to $k\sim 0.5$ $h$/Mpc and of the initial power spectrum to $k\sim 0.4$ $h$/Mpc in certain cases at $z=0$.
\end{abstract}

\begin{keywords}
cosmology: theory -- large-scale structure of Universe
\end{keywords}

\section{Introduction}
\label{sec:intro}

Local density transforms, such as the log and Gaussianization transform \citep{NeyrinckEtal2009, Neyrinck2011, neyrinck2011b}, the log transform modified for a Poisson-sampled field, known as the $A^*$-transform \citep{carronAstar, wolk2015a, wolk2015b}, and clipping \citep{simpsonClipping, simpson2015} have been proposed recently as methods to efficiently extract cosmological information from galaxy clustering data. The Gaussianization transform has been shown to restore small-scale information in the 2-point statistics of the matter density field by reducing the covariance on small scales and providing better fidelity to the linear-theory shape, and thus tightening constraints on cosmological parameters \citep{NeyrinckEtal2009, Neyrinck2011}. It has also been suggested that Gaussianization may have the ability to separate galaxy bias from underlying clustering statistics \citep{NeyrinckEtal2014, neyrinck2014cp}. Under the assumption that the bias of different fields (red and blue galaxies, dark matter) is encoded in the different 1-point probability density functions (PDFs) of the fields, and the the underlying clustering of the fields is the same, the Gaussianized fields of the biased tracers and the dark matter will have the same 2-point statistics.

While both the $A^*$-transform and clipping have been applied to real galaxy data \citep{wolk2015a, simpson2015}, previous work on the Gaussianization transform for the power spectrum has focused largely on theoretical motivation and idealized cases such as the dark matter distribution in $N$-body simulations. It is not known to what extent the benefits of the transform can be achieved in realistic galaxy samples. The effects of number density, shot noise, clustering properties, redshift-space distortions, and size of the density grid on the effectiveness of the Gaussianization transform have not been thoroughly explored.

In this paper, we apply the Gaussianization transform to realistic galaxy samples, produced from the semi-analytic galaxy formation code  {\scshape galform} \citep{cole2000, baugh2005, bower2006, galform2014, lacey2015}, with the aim of testing the ability of the transform to remove the effect of nonlinear (scale-dependent) galaxy bias to recover the small-scale shape of the matter power spectrum. We apply the transform to blue and red galaxies from {\scshape galform} \citep{galform2014} and compare the shape of the power spectra of the transformed fields to each other as well as to those of the underlying dark matter and initial density fields. We study how the transform performs in real and redshift space, and explore the effects of varying the sample selection and the size of the density grid.

In Section \ref{sec:sec1} we describe the simulations and galaxy samples used in this work. In Section \ref{sec:sec2} we discuss the Gaussianization transform in detail and present the methods used to estimate the Gaussianized power spectrum, including a prescription for shot noise correction. In Section \ref{sec:results} we present our results comparing the Gaussianized and usual density statistics of galaxies and discuss the effects of sample selection and varying cell size. In Section \ref{sec:clipping}, we compare the Gaussianization transform to clipping of the galaxy density field as a method for removing nonlinearity and nonlinear bias. We conclude in Section \ref{sec:conclusion}.

\section{Simulation and Galaxy Samples}
\label{sec:sec1}

We use simulated galaxy samples from the {\scshape galform} semi-analytic galaxy formation model \citep{galform2014}, which was run on the Millennium dark-matter-only $N$-body simulation \citep{springel2005, guo2013}\footnote{Data from the Millennium simulation is available on a relational database accessible from \url{http://galaxy-catalogue.dur.ac.uk:8080/Millennium}}.  Semi-analytic galaxy formation models such as {\scshape galform} model the formation and evolution of galaxies using simple, physically motivated equations to predict baryonic physics within dark matter halo merger trees  \citep{cole2000, baugh2005, bower2006, lacey2015}. The output is a galaxy catalogue with realistic clustering. Such a catalogue is sufficient for our purposes because it allows us to apply the Gaussianization transform to realistic galaxy samples with different clustering properties, and to test its effectiveness in recovering the known shape of the underlying dark matter and linear power spectra. 

The dark-matter only WMAP7 Millennium simulation (MR7) uses a periodic box with $L_{\text{box}}=500$ Mpc/$h$ on a side, with $2160^3$ dark matter particles, and WMAP7 cosmological parameters: $\Omega_{\textrm m}=0.272$, $\Omega_{\textrm b}=0.0455$, $\Omega_{\Lambda}=0.728$, $h=0.704$, $n_{\textrm s}=0.967$, and $\sigma_8=0.810$ \citep{guo2013, wmap7}. We use the galaxy and dark matter distributions at $z=0$.

\begin{figure}
\centering
\includegraphics[width=0.48\textwidth]{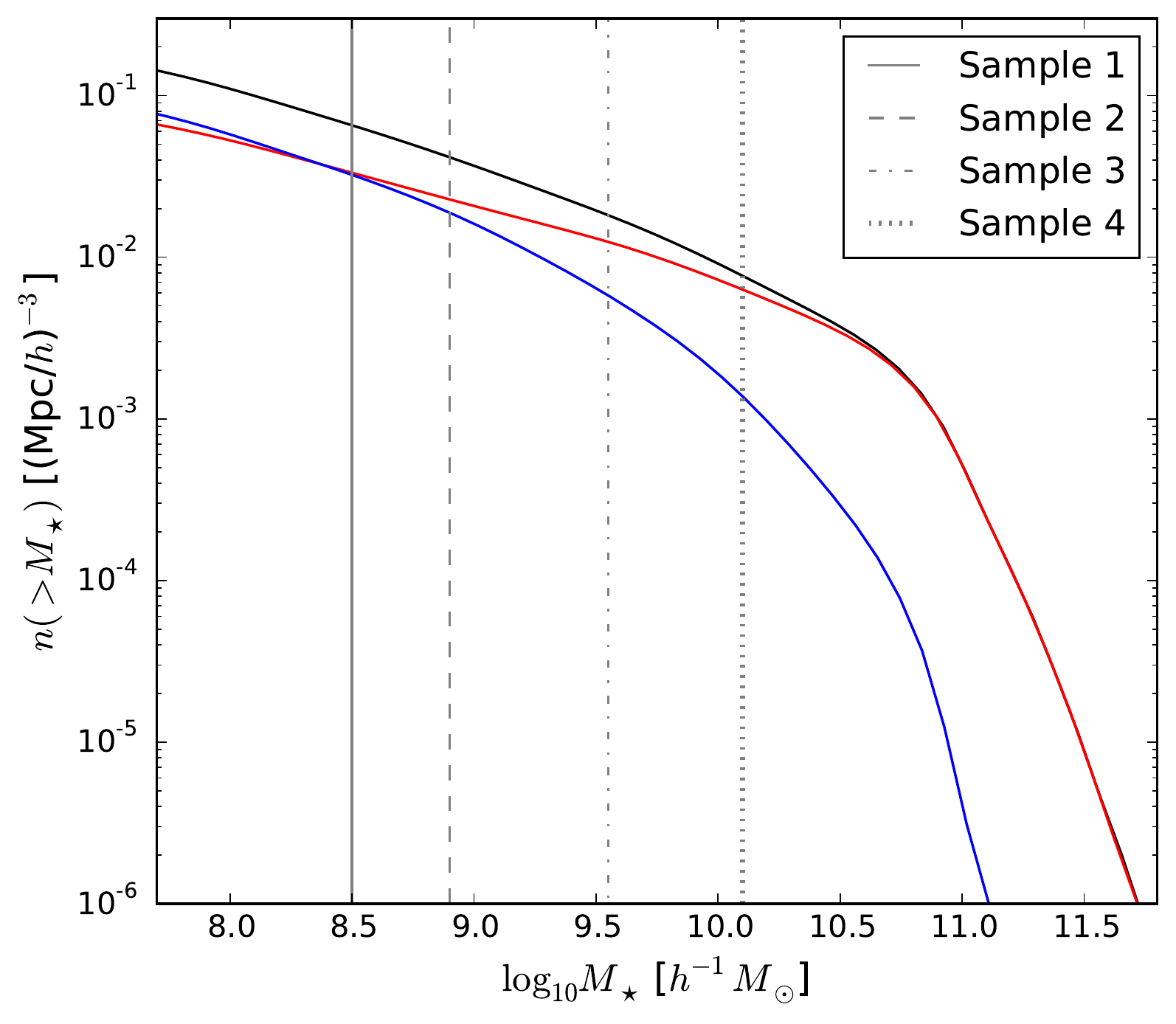}
\caption{Cumulative stellar mass function of the {\scshape galform} galaxies. The red and blue lines refer to the stellar mass functions of the red and blue galaxies, and the black solid line is the total. We make cuts in stellar mass to form 4 different samples, which are shown by the grey solid, dashed, dot-dashed, and dotted vertical lines.}
\label{fig:smf}
\end{figure}

\begin{figure}
\begin{center}
\includegraphics[width=0.48\textwidth]{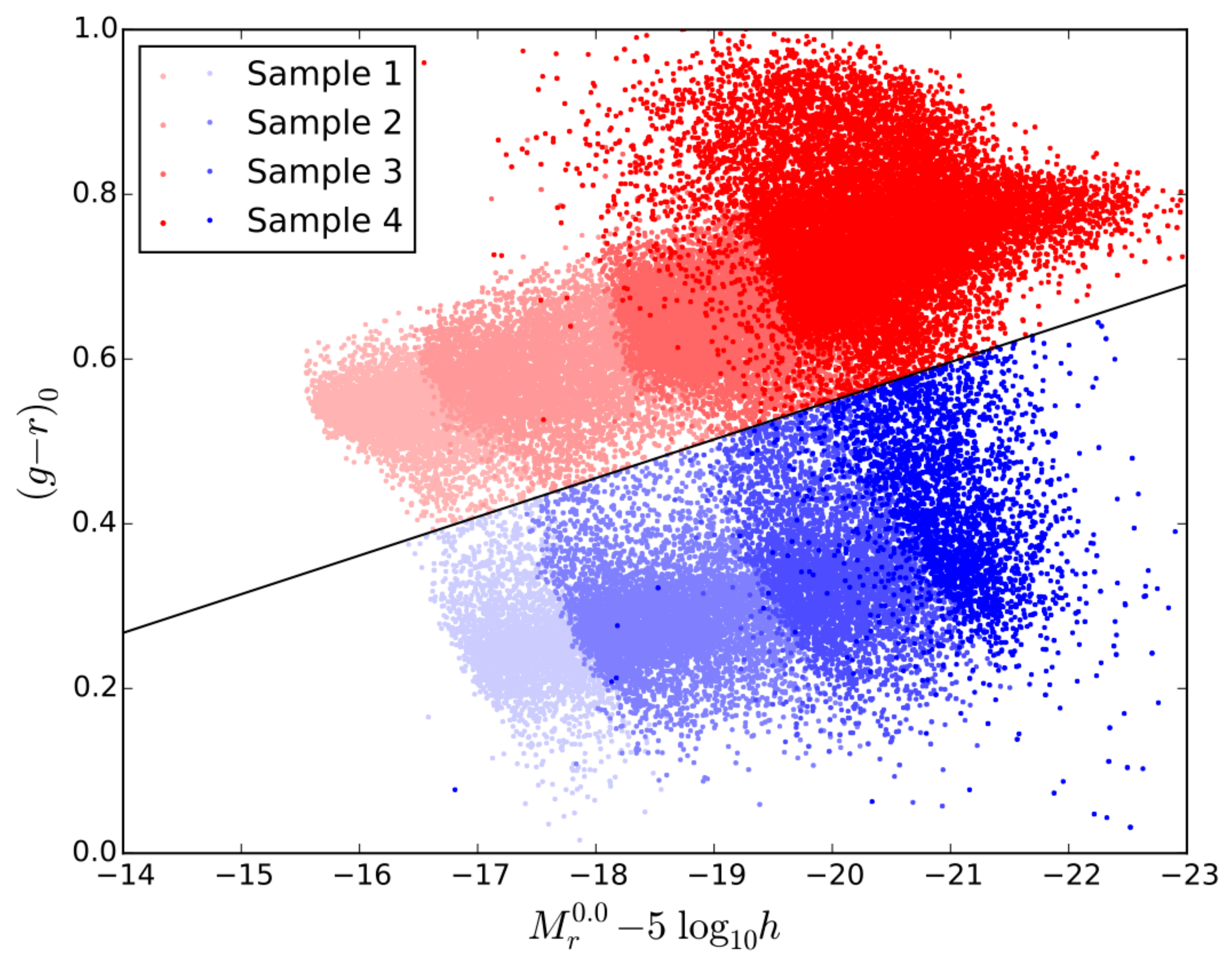}
\caption{Colour-magnitude diagram for our 4 samples. The solid line shows the colour cut used to separate red (above the line) and blue (below the line) galaxies.}
\label{fig:gal_col}
\end{center}
\end{figure}

Fig. \ref{fig:smf} shows the total stellar mass function of the {\scshape galform} galaxies (black solid line), as well as that of the red and blue galaxies separately. We create four samples using stellar-mass cuts of $10^{8.5}$, $10^{8.9}$, $10^{9.6}$ and $10^{10.1}$ $h^{-1} \textrm{M}_{\odot}$. The number densities, fraction of red galaxies, and other properties of the four samples are summarized in Table \ref{tab:samples}. We use the colour-magnitude cut shown in Fig. \ref{fig:gal_col} (solid black line) to split the galaxies into red and blue. The equation of the colour cut is: $(g-r)_0=a(M_r^{0.0}+20)+b$, with $a=-0.047$ and $b=0.55$. The colour-magnitude distributions of each sample are shown in different shades of red and blue. 

\begin{figure*}
\centering
\includegraphics[width=1.0\textwidth]{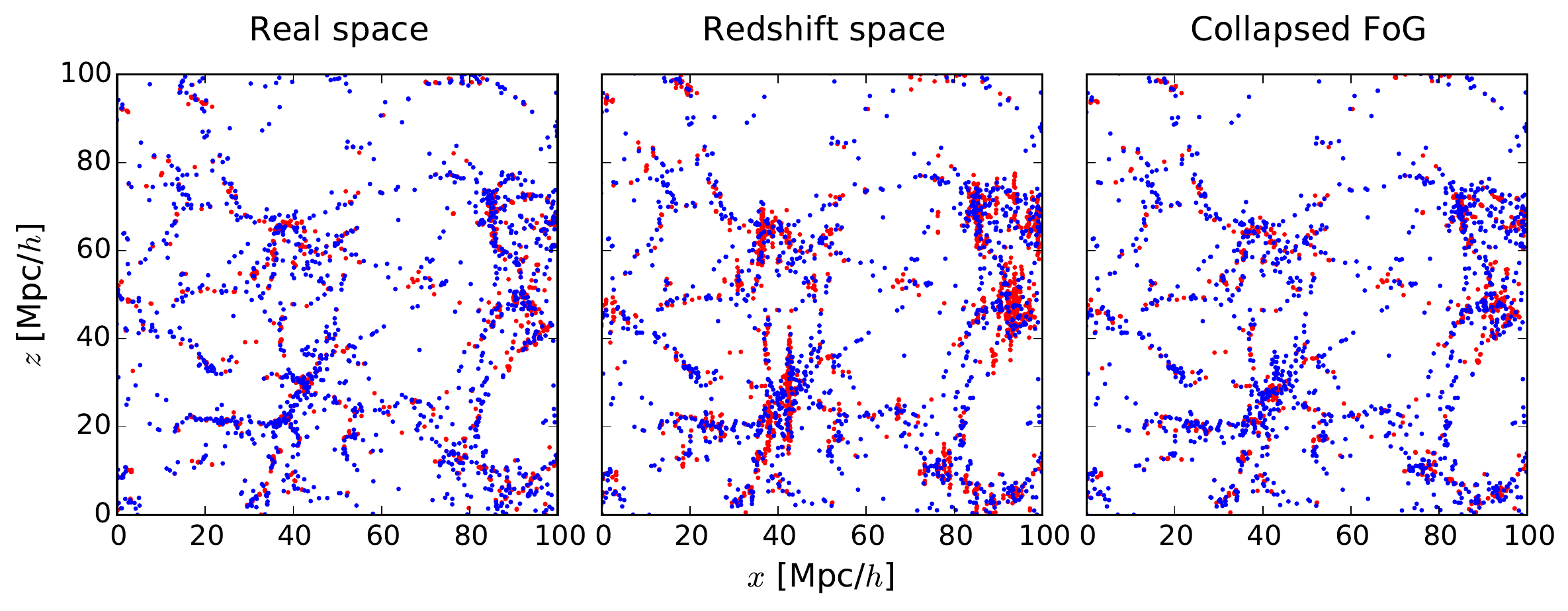}
\caption{Spatial distribution of red and blue galaxies (with corresponding colour dots) in a $7$ Mpc/$h$ slice from Sample 2 in real space (left), redshift space (centre), and with collapsed FoG (right). The $z$-axis is the line-of-sight direction and we use the distant observer approximation for the redshift-space distribution. }
\label{fig:gal_pos}
\end{figure*}

In general, red galaxies are much more strongly clustered than blue galaxies (e.g. \citet{norberg2002, zehavi2002, zehavi2011}), and this can also be seen in the {\scshape galform} galaxies \citep{campbellgalform, farrow2015}.
The left panel of Fig. \ref{fig:gal_pos} shows the positions of galaxies in a slice of the simulation box in real space. In redshift space, where the positions of the galaxies are distorted along the line of sight by their peculiar velocities, the small-scale behaviour of the two fields is very different. Red galaxies have much stronger fingers of god because they live in much more massive clusters than blue galaxies. This is shown in the middle panel of Fig. \ref{fig:gal_pos}. As we will discuss in Section \ref{sec:results}, we find it useful to also consider galaxy fields where the fingers of god have been collapsed. This is possible to do in real galaxy data with high enough number density and a reliable group catalogue \citep{tegmark2004, berlind2006}. We perform an idealized version of such a procedure here by assigning each galaxy the central subhalo velocity as opposed to its own peculiar velocity. This leads to the field in the right panel of Fig. \ref{fig:gal_pos}, which includes the large-scale distortion effect, but the fingers of god are collapsed. We refer to this as ``collapsed FoG space" throughout the paper.

\begin{table*}
\centering
\caption{Summary of galaxy samples used. The columns give the minimum stellar mass, total number density, fraction of red galaxies ($f_R$), fraction of empty cells in a (real space) $128^3$ CIC grid for red and blue galaxy samples, and the average number of galaxies of a given type per halo with a galaxy of that type.}
\label{tab:samples}
\begin{tabular}{r l c c c c c c c}
&$\textrm{M}_{\star}^{\textrm{min}}$ &$\bar n$& $f_R$ & \multicolumn{2}{|c|}{$f^{128}_{\text{empty}}$}  &   \multicolumn{3}{|c|}{$\langle N_{\text{gal}}\rangle$}   \\
&[$\textrm{M}_{\odot}/h$] &  [(Mpc/$h$)$^{-3}$]  & & Red& Blue &Red & Blue &All\\
\hline
Sample 1 &$10^{8.5}$ & 0.061 & 0.51& 0.24 & 0.05 &  3.89     &   1.08     &  1.74 \\
Sample 2 &$10^{8.9}$& 0.039&0.55 & 0.27 & 0.10  &    3.07       &   1.04  & 1.69 \\
Sample 3 &$10^{9.6}$& 0.017 & 0.69 &  0.33 & 0.30 &  2.07      &   1.02   & 1.61 \\
Sample 4 &$10^{10.1}$& 0.007 & 0.83 &  0.44 & 0.66 &  1.69     &   1.01  & 1.52
\end{tabular}
\end{table*}

\section{Gaussianization Transform and Power Spectrum Estimation}
\label{sec:sec2}

\subsection{Gaussianization Transform}
\label{sec:sec2sub1}
It has long been known that cosmological density fields are roughly lognormal \citep{ColesJones1991}. The statistics of the log-transformed density field have been shown to have lower covariance on small scales and better fidelity to the linear-theory shape \citep{NeyrinckEtal2009}. In principle, using the statistics of the log-density as opposed to the usual density variable can give much tighter constraints on cosmological parameters \citep{Neyrinck2011, neyrinck2011b}. In the case of a perfectly lognormal field, analysis of the log-density correlation function accesses all of the Fisher information, whereas using arbitrarily high-point density correlation functions can give only a small fraction of that total information on small scales \citep{Carron2011,CarronNeyrinck2012}.

Previous studies of the log transform have focused on the dark-matter density field from $N$-body simulations. The dark matter density ($\rho$) is interpolated to a grid using either a nearest-grid-point (NGP) or a cloud-in-cell (CIC) interpolation scheme \citep{hockneyeastwood}, and the statistics of $A=\ln(1+\delta)$ (where $\delta \equiv \rho / \bar \rho - 1$ is the overdensity) are analysed as opposed to those of $\delta$ itself. In reality we cannot observe the full dark matter distribution, and instead we infer it through the presence of galaxies, which act as biased tracers of the dark matter distribution. Depending on the number density and grid size, the log transform of the galaxy density field may not be well defined, as there may be many grid cells with zero density. For these cases, one option is the $A^*$-transform, providing an optimal local estimate of the log-density for a Poisson-lognormal field \citep{carronAstar}. Another is Gaussianization, which is agnostic regarding an assumed underlying continuous field and how it is point-sampled. Gaussianization maps the 1-point PDF of the density field onto an exactly Gaussian PDF with a specified variance by rank-ordering the density cells and transforming them using the inverse error function \citep{Neyrinck2011, neyrinck2011b}:
\begin{align}
\text{Gauss}(\delta)&=\sqrt 2 \sigma\ \text{erf}^{-1}\left(2f_{<\delta}-1+\frac{1}{N}\right),\label{eq:gaussxform}
\end{align}
where $f_{<\delta}$ is the fraction of cells below $\delta$, $\sigma$ is the standard deviation of the Gaussian, and $N$ is the total number of cells. The $1/N$ term in Equation \ref{eq:gaussxform} ensures that the argument in the inverse error function is between $-1$ and $1$, and thus returns a finite value for all $\delta$. For an exactly log-normal distribution, Gaussianization is equivalent to a log transform. In the case of poor sampling, where there are empty cells, all zero cells are mapped to a single value, which is set by requiring the Gaussian to have mean zero and variance $\sigma^2$. The transform thus can increase the effect of shot noise by accentuating the difference between cells with zero and nonzero density. As we will test and confirm in Section \ref{sec:results}, choosing a grid size corresponding to an average of about 1 particle per cell ensures that the contribution from shot noise is manageable and the transform delivers the greatest benefits.

\begin{figure*}
\centering
\subfloat{
\includegraphics[width=0.5\textwidth]{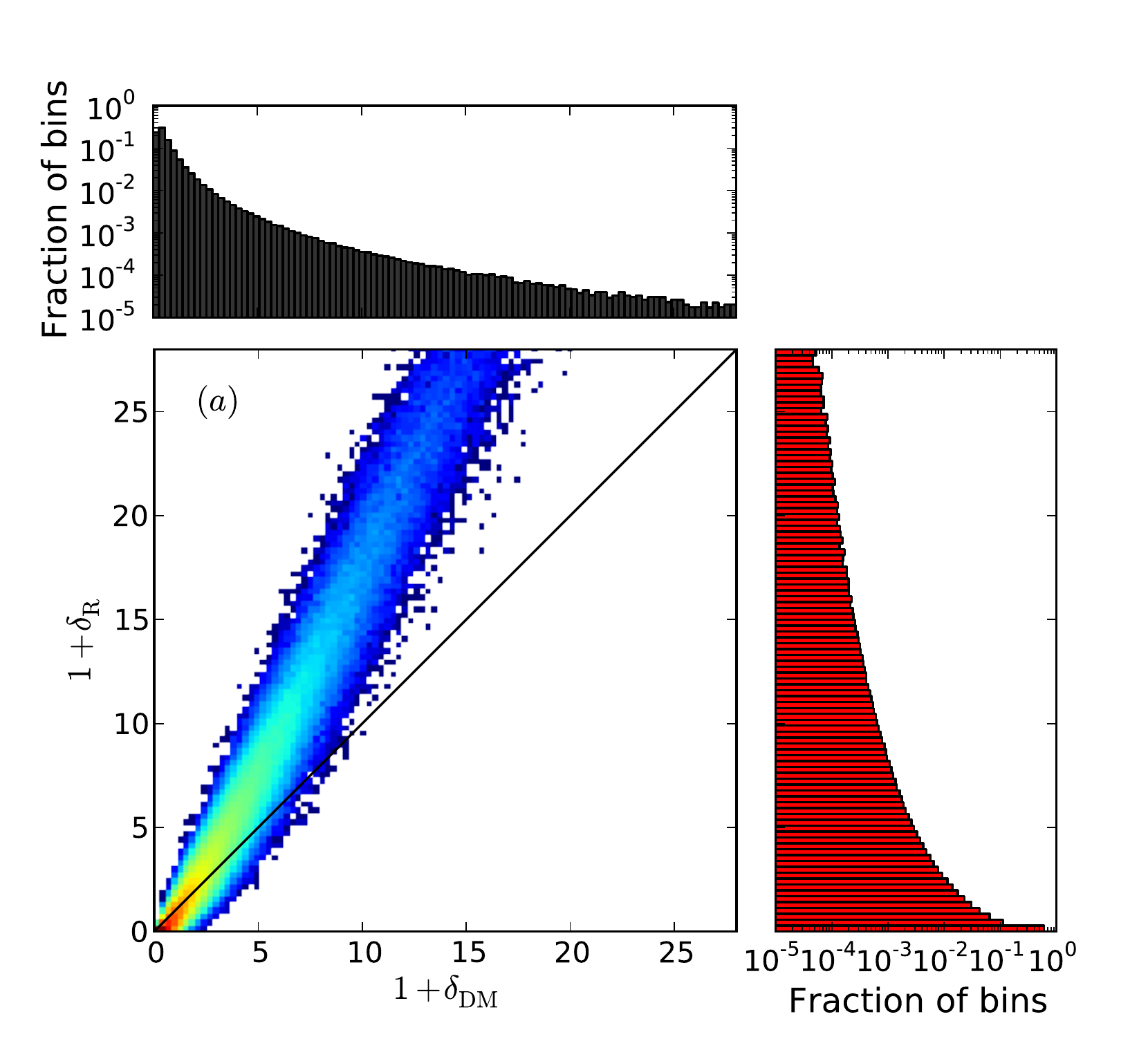}
}
\subfloat {
\includegraphics[width=0.5\textwidth]{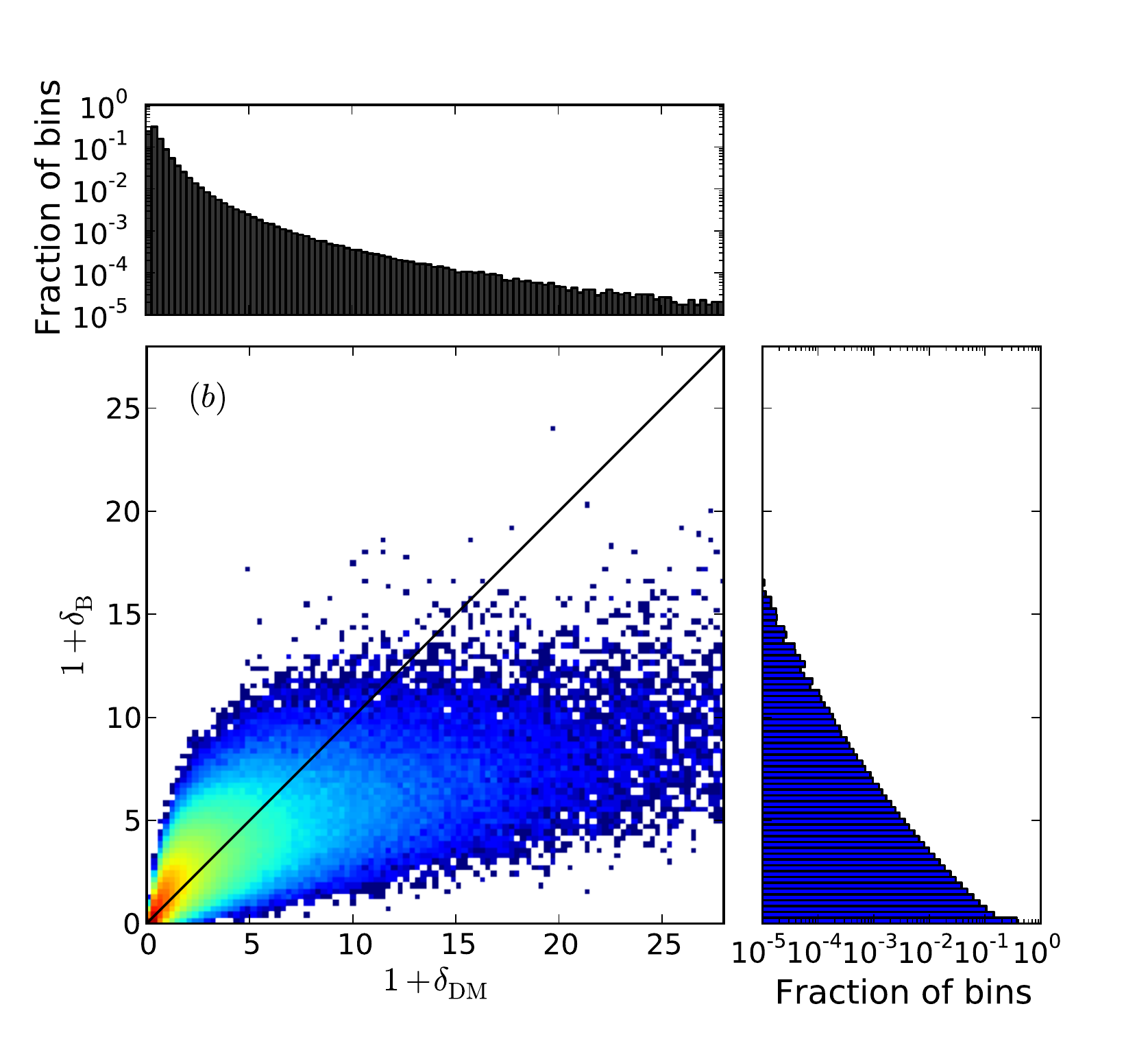}
}
\caption{Distribution of red (left panels) and blue (right panels) galaxy densities versus dark matter density on a $128^3$ grid for Sample 2 in real space. Side panels show the 1-point PDFs of the dark matter (top) and galaxies (right) and the centre panels show the 2-dimensional distribution in each case. Colours in centre panels show number of cells using a logarithmic colour mapping.}
\label{fig:gal_delta}
\end{figure*}

We explore the Gaussianization transform first through studying the relationship between the usual and transformed galaxy and dark matter density fields. We use the galaxies from Sample 2 in real space to illustrate these relationships. Fig. \ref{fig:gal_delta} shows 2-dimensional histograms of the red (a) and blue (b) galaxy densities from Sample 2 versus the dark matter density on $128^3$ CIC grids, as well as the 1-point PDFs of each field. From these figures it is clear that red galaxies are highly biased with respect to the dark matter, whereas blue galaxies have bias less than 1. The relationship between red galaxy density and dark matter density is also quite tight compared to the blue galaxies over this range of densities.

\begin{figure}
\centering
\includegraphics[width=0.5\textwidth]{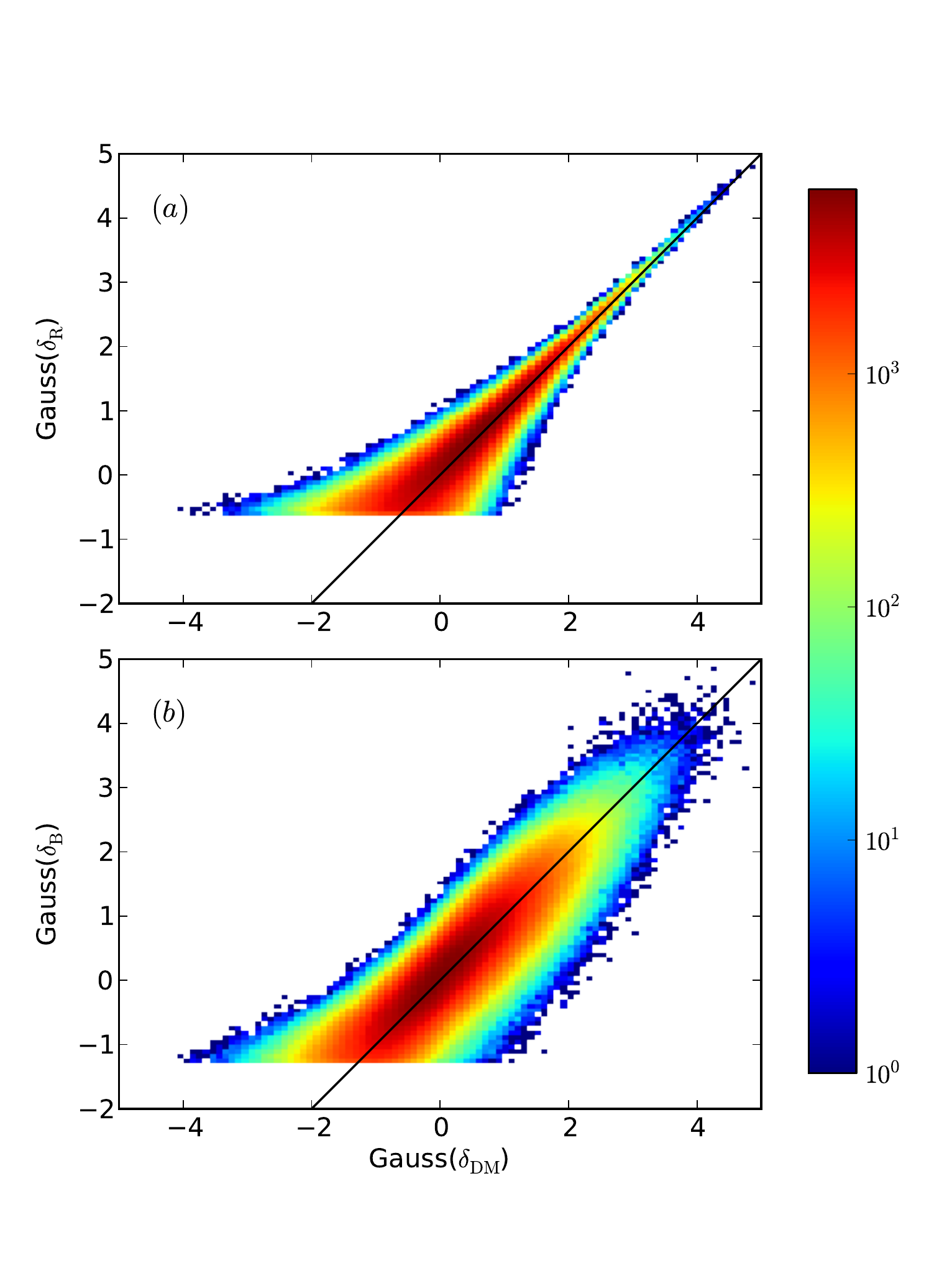}
\caption{Distribution of Gaussianized density of red (a) and blue (b) galaxies versus Gaussianized dark matter density on a $128^3$ CIC grid for Sample 2. In both panels, we show only cells where the galaxy density is non-zero and there are more empty cells for the red galaxy sample. Colour bar gives the number of cells with a logarithmic colour mapping.}
\label{fig:gauss_hist}
\end{figure}

Next, we consider the relationship between the Gaussianized galaxy fields and the Gaussianized dark matter field. Fig. \ref{fig:gauss_hist} shows the 2-dimensional histograms of the Gaussianized red (a) and blue (b) fields versus the Gaussianized dark matter field from Sample 2. Note that we only show the non-zero density cells in these figures. For the Gaussianization transform, we have used Equation \ref{eq:gaussxform} with $\sigma=1.0$ in all cases. The value of $\sigma$ used is arbitrary, and only affects the amplitude of the Gaussianized field, and thus affects the amplitude of the Gaussianized statistics but not the scale-dependence. All the zero-density cells in the galaxy fields map to a single value in the Gaussianized field. The relationship between the Gaussianized red galaxy field and dark matter is remarkably tight, especially in high-density regions. The Gaussianized fields of blue galaxies and dark matter are clearly correlated, but with more scatter in high density regions than in the case of the red galaxies. For example, if we consider cells in both the red and blue galaxy fields that have Gaussianized density values close to 3.0 ($\pm 0.05$), the standard deviation of the corresponding Gaussianized dark matter density cells is much smaller for the red (0.08) than the blue (0.41).

\begin{figure}
\centering
\includegraphics[width=0.5\textwidth]{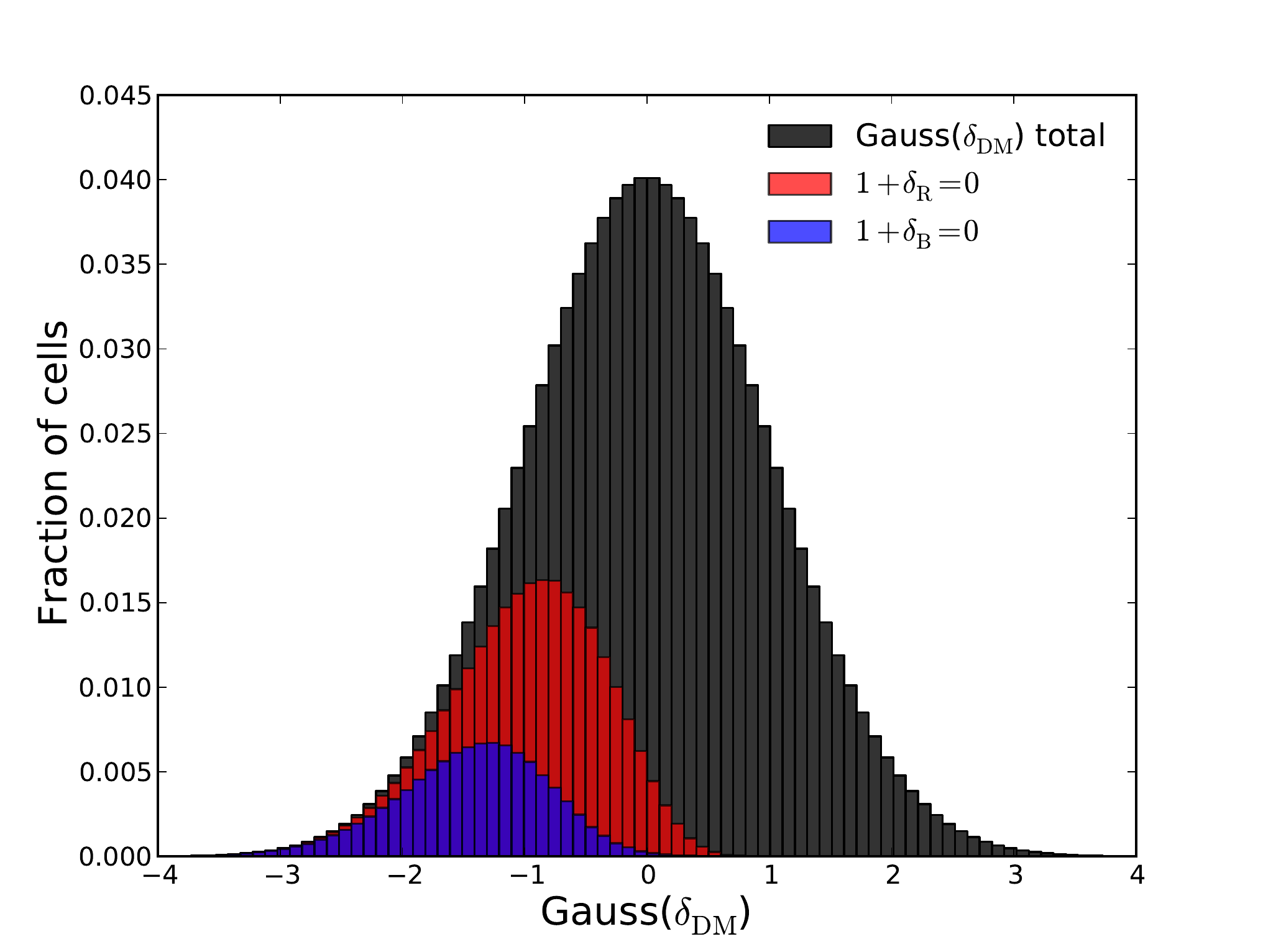}
\caption{Distribution of Gaussianized dark matter density cells. In black, all cells; in red (blue), cells that are empty in the red (blue) galaxy density field for Sample 2.}
\label{fig:rb0_hist}
\end{figure}

Fig. \ref{fig:rb0_hist} shows the Gaussianized dark matter PDF, which is by definition a Gaussian with mean zero and $\sigma=1.0$. The red and blue PDFs show the distributions of dark matter cells that have zero density ($\delta=-1$) in the red and blue galaxy fields, respectively. For the CIC density, zero-density cells correspond to cells with no particles and whose neighbouring cells also have no particles. Note that although the galaxy fields have similar number densities in this sample ($\bar n_R=0.021$ and $\bar n_B = 0.018$ (Mpc/$h$)$^{-3}$), there are more empty cells in the red galaxy field because it is more highly clustered than the blue.

\subsection{Gaussianized Statistics}
\label{sec:sec2sub2}

Next, we discuss methods used to estimate the power spectra of the Gaussianized fields, normalise the large-scale amplitude of the power spectra, and correct for shot noise. 

Raw power spectra are computed through an FFT of the field under consideration (the $\delta$ field or the transformed field). We square the Fourier amplitudes to obtain the 3-dimensional power spectrum of the field, and then average over shells in $|\vec k |$ to find $P(k)$. In redshift space, this computes the angular-average or monopole of the full 2-dimensional redshift-space power spectrum.

The large-scale amplitude of a Gaussianized power spectrum depends on various quantities, including the intrinsic amplitude of fluctuations, the large-scale bias, the variance of the Gaussian, and the grid size used. While this amplitude does contain information, in this work we focus only on the shape of the power spectrum, and thus we choose to normalise all galaxy power spectra to the large-scale amplitude of the real-space dark matter power spectrum at $z=0$. We define the large-scale amplitude of a given power spectrum as:
\begin{align}
B=\frac{\sum_i P(k_i) N_i}{\sum_i N_i},\label{eq:lsamp}
\end{align}
where $N_i$ is the number of modes in bin $i$, and the sum is taken for $k<0.07$ $h$/Mpc, to include only linear modes. For example, to normalize a Gaussianized power spectrum $P_{\text{Gauss}}$ to the dark matter power spectrum $P_{\text{DM}}$, we compute $B$ for each and multiply the Gaussianized power spectrum by $B_{\text{DM}}/B_{\text{Gauss}}$.

We must also correct for the effects of shot noise in our power spectra. For the usual $\delta$ power spectrum, we correct for Poisson shot noise by subtracting $1/\bar n$, where $\bar n$ is the tracer number density. For the power spectra of the Gaussianized fields, we apply a shot-noise correction procedure, detailed below.

In general, the contribution of shot noise in the Gaussianized power spectrum is more complicated than in the usual density power spectrum, and may depend on scale, number density, grid size, and intrinsic clustering properties of the sample. However, \citet{Neyrinck2011} found that the contribution from shot noise in the Gaussianized power spectrum does not diverge greatly from a constant form on quasilinear scales. Therefore, we make the ansatz that on these scales, the shot noise will contribute in a scale-independent way:
\begin{align}
P_{\text{meas}}^{\bar n}(k)&=P_{\text{true}}(k)+\frac{A}{\bar n} \label{eq:shotnoise},
\end{align}
where $A$ is an unknown amplitude that we must fit for in a given galaxy sample. In general, $A$ may depend on the clustering of the given sample, so its value will be different for red and blue galaxies, and will depend on the sample definition.

We test this ansatz by randomly down-sampling a given galaxy sample to several different values of $\bar n$ and computing the resulting Gaussianized power spectra. We look at the differences between these power spectra, which arise from the difference in shot noise and so should depend only on the difference between the reciprocals of the number densities, if our ansatz is valid:
\begin{align}
P_{\text{meas}}^{\bar n_1}(k)-P_{\text{meas}}^{\bar n_2}(k)&=A \left (\frac{1}{\bar n_1}-\frac{1}{\bar n_2}\right) \label{eq:sndiff}.
\end{align}

We estimate $A$ for each pair of $\bar n_1$, $\bar n_2$ using the above expression and weight by the number of modes in each $k$-bin, $N_i$:
\begin{align}
\hat A(\bar n_1, \bar n_2)&=\left( \frac{\bar n_2-\bar n_1}{\bar n_1 \bar n_2}\right)\frac{\sum_i\left(P_{\text{meas}}^{\bar n_1}(k_i)-P_{\text{meas}}^{\bar n_2}(k_i)\right) N_i}{ \sum_i N_i}\label{eq:aest},
\end{align}
where the sum is over bins $0.1 \le k  \le 0.3$ $h$/Mpc, as we expect our ansatz to apply over this range of scales.  We use 4 different sample sizes for each sample, giving 6 combinations of pairs of ($\bar n_1$, $\bar n_2$), and we weight all pairs equally to find the best-fit value of $A$ to correct for the shot noise in Equation \ref{eq:shotnoise}.

\begin{figure} 
\centering
\includegraphics[width=0.485\textwidth]{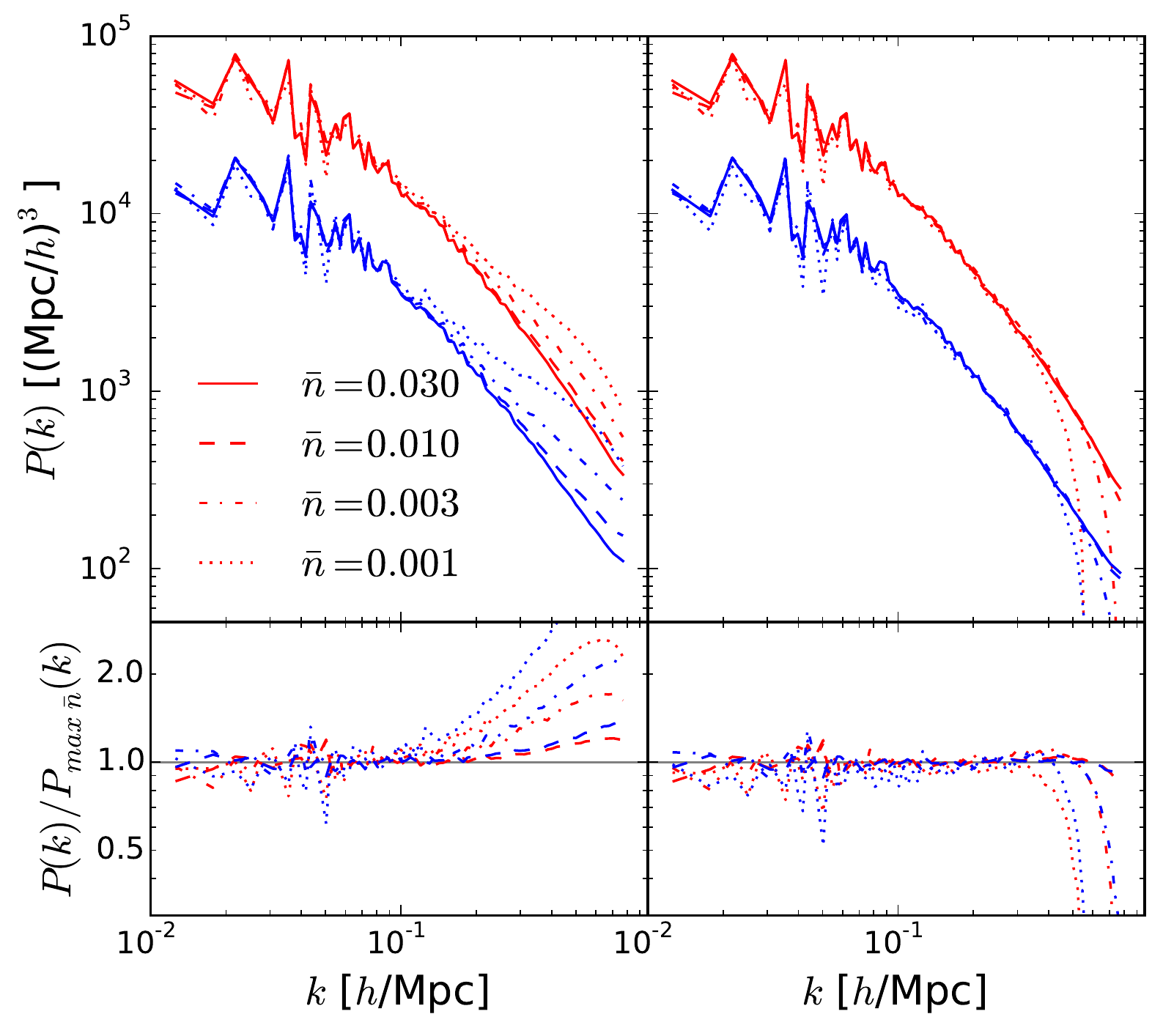}
\caption{Gaussianized power spectra with fixed large-scale amplitudes, before (left) and after (right) shot-noise correction. The lower panels show the ratios of the down-sampled power spectra to that of the largest sample ($\bar n=0.030$). The deviation from unity in the lower-left panel is the shot-noise term. The ratios in the lower right panel agree much better with each other, which indicates that our model for shot noise is reasonable on these scales. }
\label{fig:pksn}
\end{figure}

Fig. \ref{fig:pksn} shows the Gaussianized power spectra of the red and blue galaxies from Sample 1, down-sampled to different values of $\bar n$. The red lines show the Gaussianized power spectra of red galaxies at different number densities, and the blue lines show the power spectra of blue galaxies at the same number densities. The left panels show the power spectra before shot noise correction, and the right panels show the shot-noise-corrected power spectra. In these plots, the large-scale amplitudes of each have been normalised to that of the usual density power spectra of the galaxies of each colour using Equation \ref{eq:lsamp}. In the left panel the effect of shot noise can be seen in the high-$k$ tail of the power spectra as the sample becomes smaller. The lower left panel shows the ratio of the power spectra of the sampled galaxy population to that of the densest sample ($\bar n=0.030$ (Mpc/$h$)$^{-3}$). The deviation grows with decreasing $\bar n$. The differences between these power spectra are used to fit for the value of $A$ as in Equation \ref{eq:aest}. The upper right panel shows the power spectra after the shot noise correction has been applied, and the lower right panel shows the corresponding ratios. Note that the range of $k$ that this shot noise correction works well on depends on the number density $\bar n$.

From this figure, we see that after shot-noise correction, the power spectra converge as $\bar n$ increases. We conclude that for $\bar n\ge 0.010$ (Mpc/$h$)$^{-3}$, the procedure results in the correct power spectrum (recovers ``$P_{\mathrm{true}}(k)$'' to within 10\%) on scales larger than the Nyquist frequency $k \le 0.8$ $h$/Mpc. For $\bar n=0.003$ (Mpc/$h$)$^{-3}$ the correction is accurate up to $k\sim 0.6$ $h$/Mpc. For the lowest number density shown here, $\bar n=0.001$ (Mpc/$h$)$^{-3}$, the shot noise correction is especially poor even on relatively large scales. In general, we find that the accuracy of the shot noise procedure depends on several variables, including number density and fraction of empty cells. For the $128^3$ grid size, we limit our analysis to samples with number densities $\bar n>0.001$ (Mpc/$h$)$^{-3}$. This excludes the blue galaxy sample from Sample 4 from our analysis, as it has a number density of $\bar n=0.001$ (Mpc/$h$)$^{-3}$. For each sample and cell size, we test the shot noise correction procedure to determine on what scales it can be trusted.

\section{Results}
\label{sec:results}
We first present the measured statistics from a single sample (Sample 2 in Table \ref{tab:samples}) on a $128^3$ CIC grid. We choose this grid size because it corresponds to roughly 1 particle per cell for this sample. In Section \ref{sec:resultssub1}, we show the comparison between the galaxy and dark matter statistics for both the usual and transformed galaxy density fields. In Section \ref{sec:resultssub2}, we compare the usual and transformed statistics to that of the linear power spectrum, and in Section \ref{sec:resultssub3} we compare the statistics of the red galaxy field to those of the blue. In Section \ref{sec:resultssub4}, we discuss the effects of varying both the sample selection and the grid size on the results.

\subsection{Galaxy versus Dark Matter Statistics}
\label{sec:resultssub1}

\begin{figure}
\centering
\includegraphics[width=0.485\textwidth]{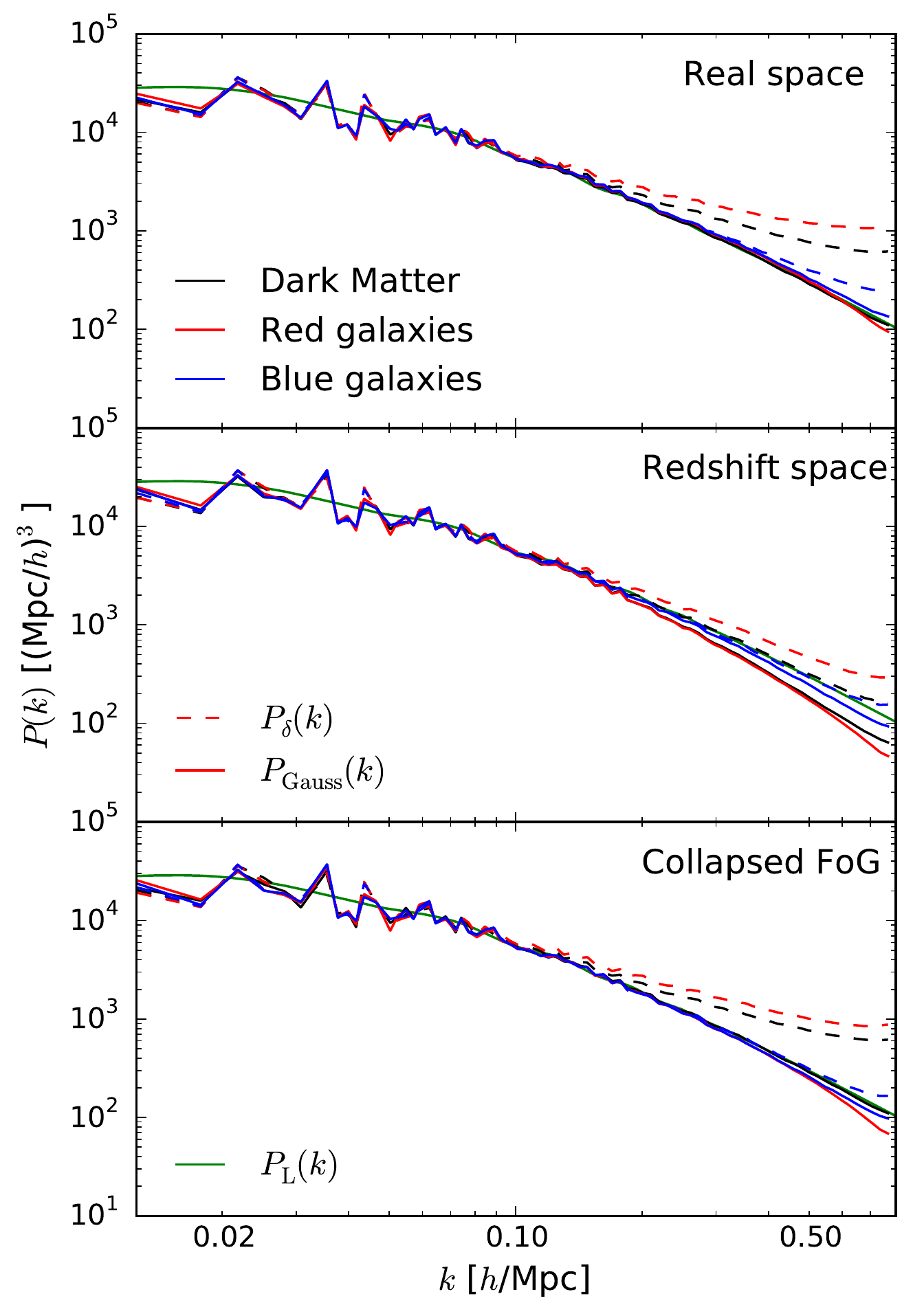}
\caption{Power spectra of dark matter (black lines), red galaxies (red lines), and blue galaxies (blue lines) in real space (top panel), redshift space (middle panel), and collapsed FoG space (bottom panel). Solid lines show Gaussianized density power spectra and the dashed lines show the usual density power spectra. The green line (identical in all panels) shows the linear (input) power spectrum at $z=0$. Large-scale amplitudes of all $P(k)$ are normalised to the real-space dark matter power spectrum, as described in Section \ref{sec:sec2sub2}.}
\label{fig:pkall}
\end{figure}

Fig. \ref{fig:pkall} shows the normalised, shot-noise corrected power spectra in real space (top), redshift space (middle), and with collapsed fingers of god (bottom). In redshift space and with collapsed fingers of god, we measure the angular-averaged (monopole) power spectrum. The solid black, red, and blue lines show the Gaussianized dark matter, red galaxy, and blue galaxy power spectra, respectively. The dashed lines show the corresponding power spectra of the usual density fields, and the green line shows the linear (input) power spectrum computed using {\scshape camb} \citep{camb}. In real space, the usual density power spectra (dashed lines) of the red and blue galaxies deviate significantly from the dark matter power spectrum for $k\gtrsim0.15$ $h$/Mpc. This is due to nonlinear galaxy bias. The Gaussianized power spectra (solid lines) lie more or less on top of one another to the smallest scales shown here (to the Nyquist frequency of the $128^3$ grid, $k_{\text{Ny}}=0.8$ $h$/Mpc), which supports the hypothesis that the transform removes nonlinear bias. The Gaussianized power spectra are also much closer to the linear power spectrum on small scales in real space.

\begin{figure}
\centering
\includegraphics[width=0.485\textwidth]{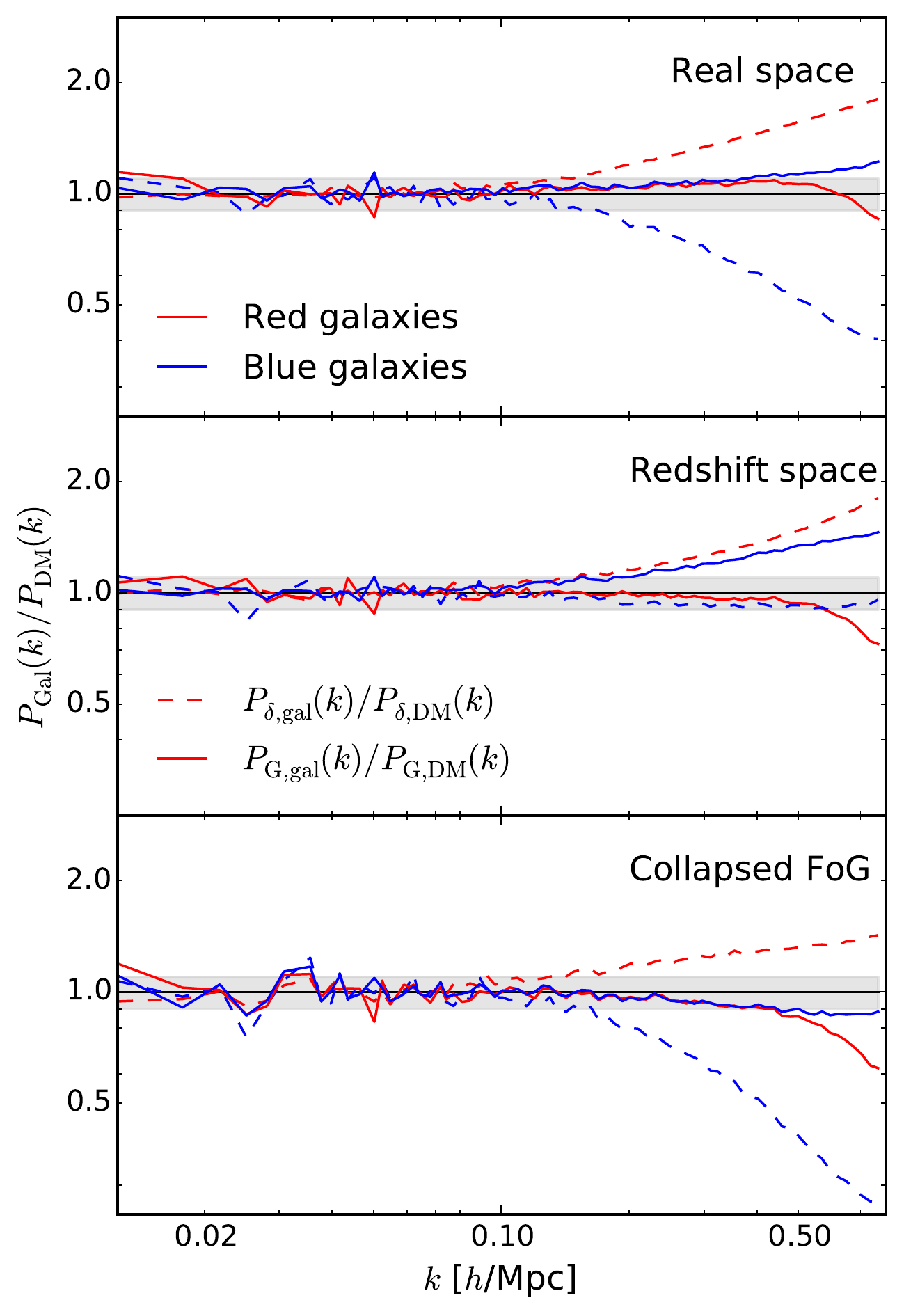}
\caption{Ratio of red and blue galaxy power spectra with dark matter power spectra in real space (top panel), redshift space (middle panel), and collapsed FoG space (bottom panel). For real space and collapsed FoG space, we compare to the real-space dark matter statistics. For redshift space, we compare to the redshift-space dark matter statistics. As in Fig. \ref{fig:pkall}, solid lines show Gaussianized power spectra and dashed lines show the usual density power spectra. Red and blue lines corresponding to red and blue galaxy statistics. Gaussianization works particularly well in redshift space at recovering the underlying Gaussianized dark matter density field.}
\label{fig:pkratioDM}
\end{figure}

The agreement with dark matter can be seen more clearly in Fig. \ref{fig:pkratioDM}, which shows the ratio of the galaxy power spectra with the corresponding dark matter power spectrum. The dashed lines show the ratio of the usual density power spectra of the galaxy fields to that of the dark matter field. The solid lines show the ratio of the Gaussianized power spectra of galaxies to the Gaussianized dark matter power spectrum. The shaded region shows 10\% deviation from the dark matter spectra. In the top and bottom panels, we compare galaxy statistics to real-space dark matter statistics. In the middle panel, we compare redshift-space galaxy statistics to redshift-space dark matter statistics.

In real space, the Gaussianized galaxy statistics have the same shape as that of the dark matter to much smaller scales ($k\sim0.4$ $h$/Mpc at 10\% level) than the usual density statistics ($k\sim0.15$ $h$/Mpc at 10\% level). However, the situation is more complicated in redshift space, where the Gaussianized red galaxy power spectrum agrees with the underlying dark matter statistics to even smaller scales than in real space ($k\sim0.6$ $h$/Mpc) but the Gaussianized blue power spectrum deviates at larger scales ($k\sim 0.2$ $h$/Mpc). Collapsing the fingers of god brings the agreement back to similar levels as in real space.

This comparison suggests that the hypothesis that Gaussianization removes nonlinear galaxy bias holds only in real space on quasilinear scales ($k\lesssim0.4$ $h$/Mpc), and that it breaks down in redshift space for the blue galaxies. With traditional density statistics, the agreement with the underlying dark matter is reversed between red and blue, with blue galaxies likely to be better described in redshift space by dark matter.

\subsection{Galaxy versus Initial Statistics}
\label{sec:resultssub2}

\begin{figure}
\centering
\includegraphics[width=0.485\textwidth]{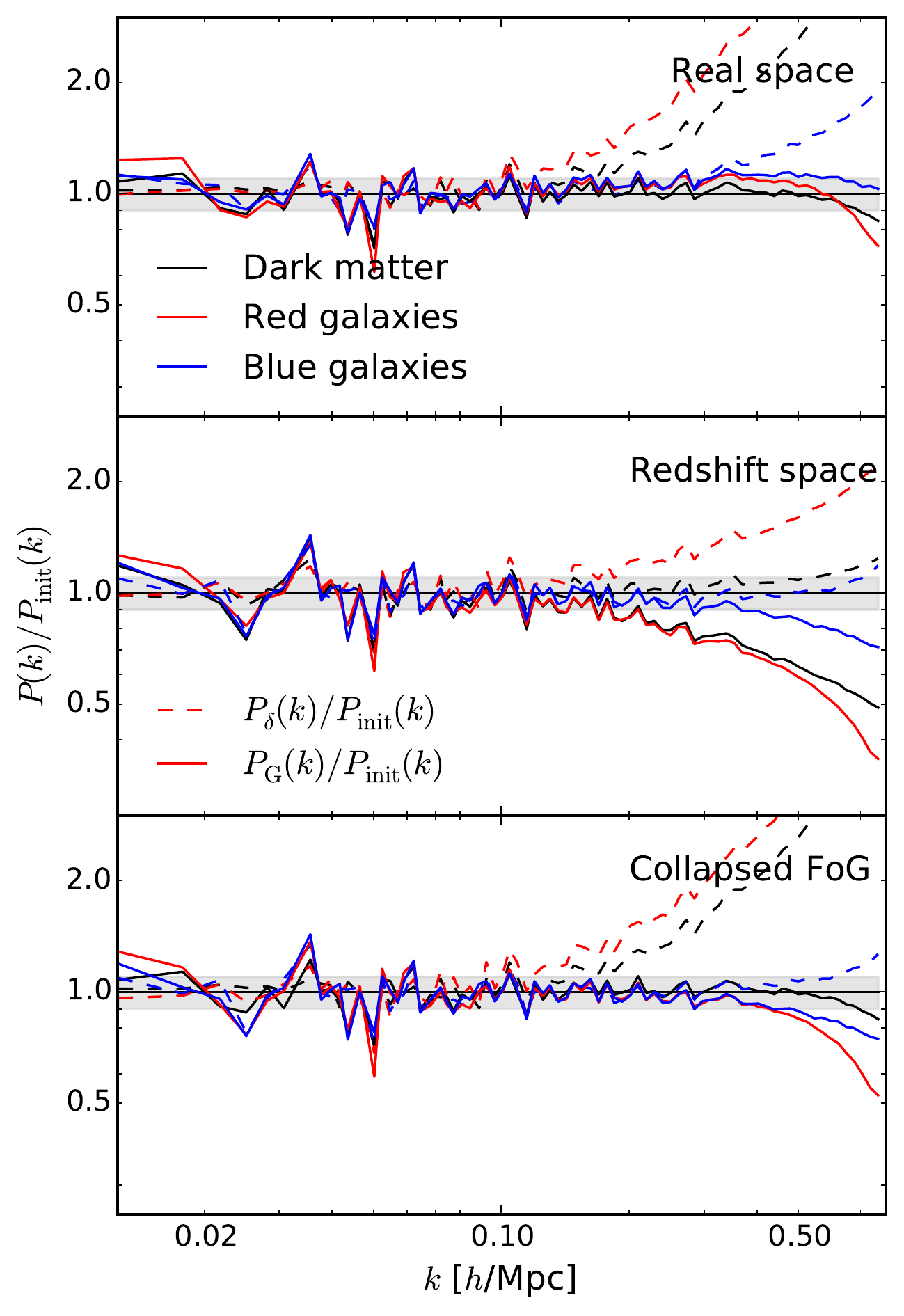}
\caption{Ratio of red and blue galaxy power spectra with the initial (linear) power spectrum in real space, redshift space, and collapsed FoG space. The denominator in all 3 panels is the same. Line colours and styles are the same as in Fig. \ref{fig:pkall}. Power spectrum of the Gaussianized galaxy density field recovers well the initial power spectrum in real and collapsed FoG space.}
\label{fig:pkratioinit}
\end{figure}

One of the original proposed goals of the log transform and Gaussianization was to reconstruct the initial power spectrum from the final density field \citep{weinberg1992, NeyrinckEtal2009}. We test this by comparing the usual and transformed density statistics to the initial power spectrum. We can see already in Fig. \ref{fig:pkall} that the real-space shapes of the Gaussianized spectra are much closer to the linear power spectrum than the usual density statistics. Fig. \ref{fig:pkratioinit} shows the ratio of the galaxy and dark matter power spectra with the dark matter power spectrum measured from an early snapshot of the Millennium simulation in real space. We use the measured initial power spectrum because it has the same shape as the linear (input) power spectrum on small scales, but includes the random modes from the simulation on large scales. The line styles and colours are the same as in Fig. \ref{fig:pkall}.

In real space, the Gaussianized statistics retain the shape of the initial power spectrum better than the usual density power spectrum. The shape of the Gaussianized dark matter power spectrum agrees with the initial power spectrum to $k\sim 0.7$ $h$/Mpc, as compared to $k\sim 0.13$ $h$/Mpc for the usual density power spectrum. The shape of the Gaussianized galaxy power spectra agree with the initial power spectrum to about $k\sim0.4$ $h$/Mpc, whereas the usual density power spectra agree to $k\sim0.1$ $h$/Mpc (red) and $k\sim0.3$ $h$/Mpc (blue).

However, the agreement again breaks down in redshift space. Interestingly, the usual density power spectrum of dark matter and blue galaxies agree with the initial power spectrum to smaller scales ($k\sim 0.5$ $h$/Mpc) than they do in real space. The Gaussianized power spectra only agree with the initial power spectrum to $k\sim 0.15$ $h$/Mpc (dark matter and red galaxies) and $k\sim0.3$ $h$/Mpc (blue galaxies). In this case, the Gaussianization transform seems not to be beneficial in recovering the small-scale shape of the initial power spectrum. However, collapsing the fingers of god restores the real-space agreement to a large extent.

This comparison is a test of how well the transform removes the effects of both nonlinear bias and nonlinear gravitational evolution from the power spectra of galaxies. Fig. \ref{fig:pkratioinit} shows that the transform is effective in achieving both goals to $k\sim 0.4$ $h$/Mpc in real space but not in redshift space. We attribute this to the fingers of god, which alter the small-scale behaviour of the galaxy and dark matter fields in different ways. Gaussianization is almost as effective at removing nonlinear bias and nonlinearity after fingers of god have been collapsed as it is in real space.

\subsection{Red versus Blue Statistics}
\label{sec:resultssub3}

\begin{figure}
\centering
\includegraphics[width=0.485\textwidth]{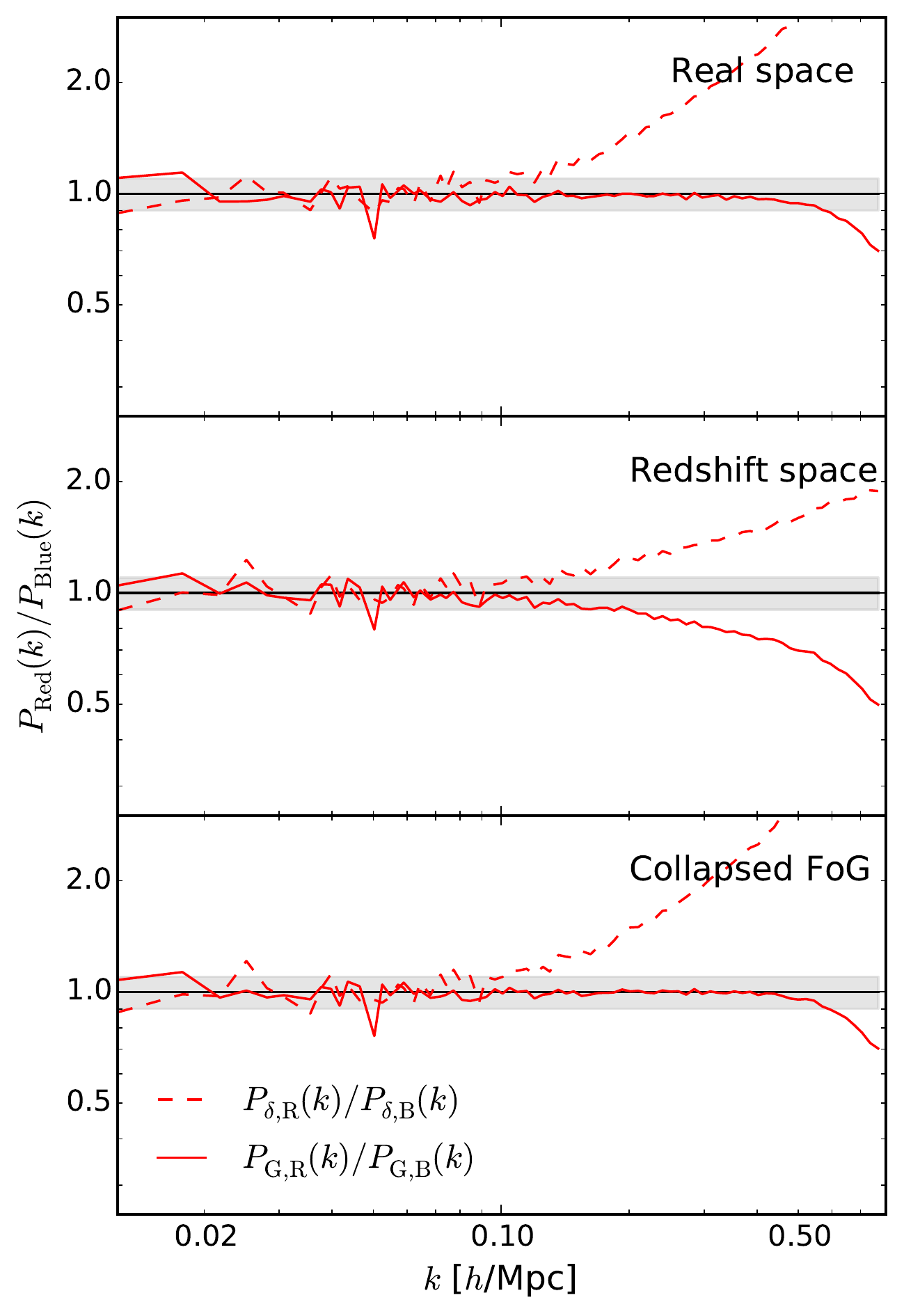}
\caption{Ratio of red to blue galaxy power spectra in real space, redshift space, and collapsed FoG space. Solid lines show the ratio of Gaussianized statistics of red and blue galaxies, and dashed lines show the ratio of the usual density statistics. The agreement between Gaussianized power spectra for the two galaxy types, once FoG are accounted for, demonstrates the potential power of Gaussianization.}
\label{fig:pkratiorvb}
\end{figure}

Because we want to test the ability of the Gaussianization transform to remove nonlinear bias, we also consider the ratio between the red and blue galaxy power spectra of both the usual and transformed fields. Fig. \ref{fig:pkratiorvb} shows this ratio in real (top), redshift (middle), and collapsed FoG (bottom) space. The dashed lines show the ratio for the usual density power spectra and the solid lines show the ratio for the Gaussianized density power spectra.

In real space, the transformed galaxy density statistics agree with each other to $k\sim 0.5$ $h$/Mpc, whereas the usual density power spectra deviate around $k\sim 0.1$ $h$/Mpc. This further supports what we observed in Fig. \ref{fig:pkratioDM}, that the statistics of Gaussianized fields of differently biased tracers agree to smaller scales than those of the usual density fields. Interestingly, it appears that the agreement between the shapes of the Gaussianized red and blue power spectra extends to smaller scales than the agreement between the galaxies and the Gaussianized dark matter (Fig \ref{fig:pkratioDM}, top panel). 

However, the agreement is lost in redshift space, presumably due to the fingers of god. In this case, the agreement of the Gaussianized statistics is only marginally better than the usual density statistics. By collapsing the fingers of god, we again see the agreement in the transformed galaxy spectra, to similar scales as in real space.

\subsection{Dependence on sample selection and cell size}
\label{sec:resultssub4}

So far we have only presented results from a single sample (Sample 2) using a cell size of $4$ Mpc/$h$, which corresponds to an average of roughly 1 particle per cell in both the red and blue samples. However, it is important to quantify the effects of varying sample selection and cell size on the results quoted above. First, we study the effect of varying the grid size in Sample 2 on the ratios considered in the previous section. Next, we look at these ratios for the 4 different samples, which have different clustering due to the stellar mass cuts, as well as different number densities. Finally, we comment on how to choose the optimal grid size for a given sample. In this section, we only show power spectrum ratios in real space, but the variations with grid size and sample are similar in redshift space and with collapsed fingers of god.

\begin{figure*}
\centering
\subfloat{
\includegraphics[width=0.485\textwidth]{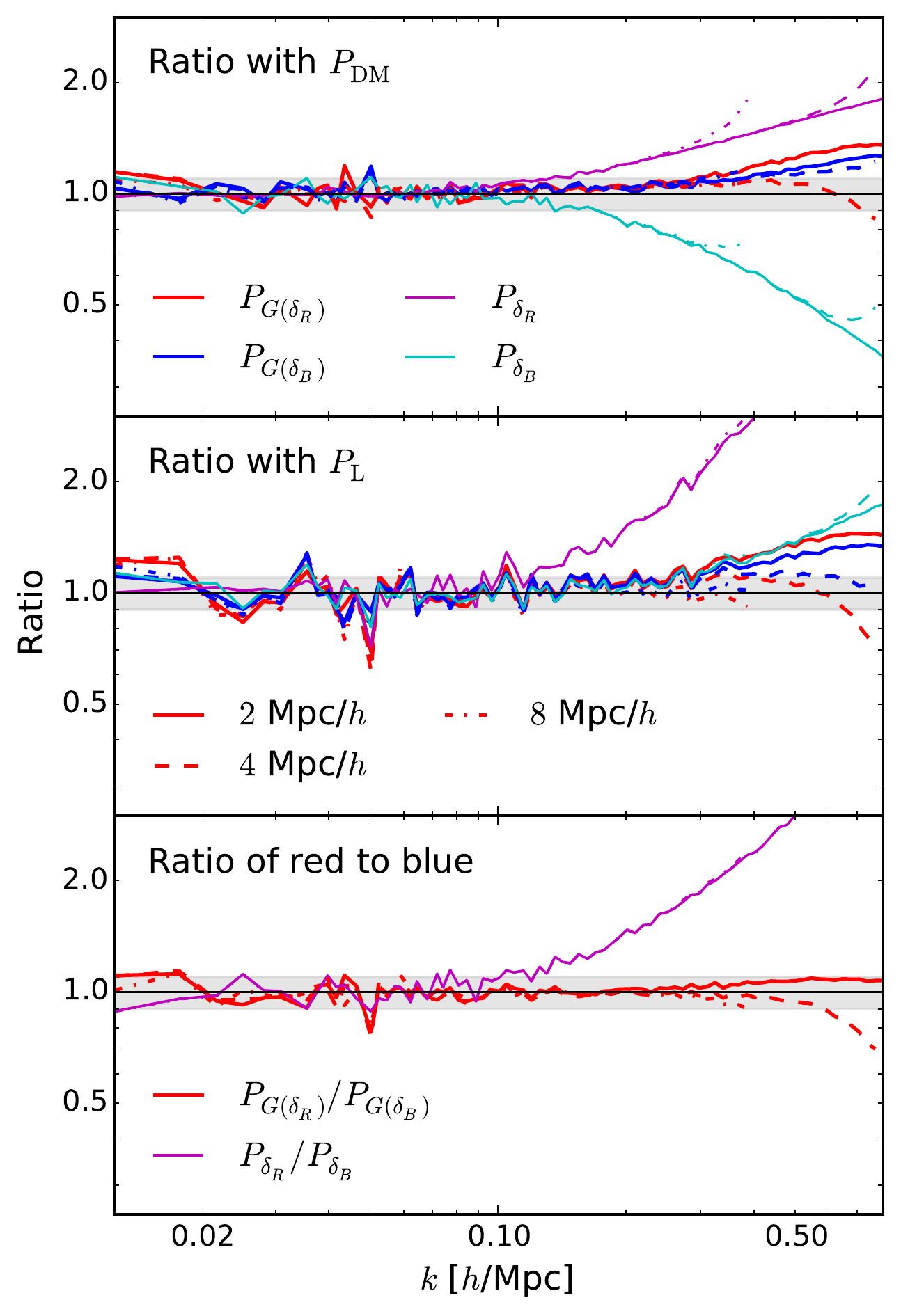}
}
\subfloat {
\includegraphics[width=0.485\textwidth]{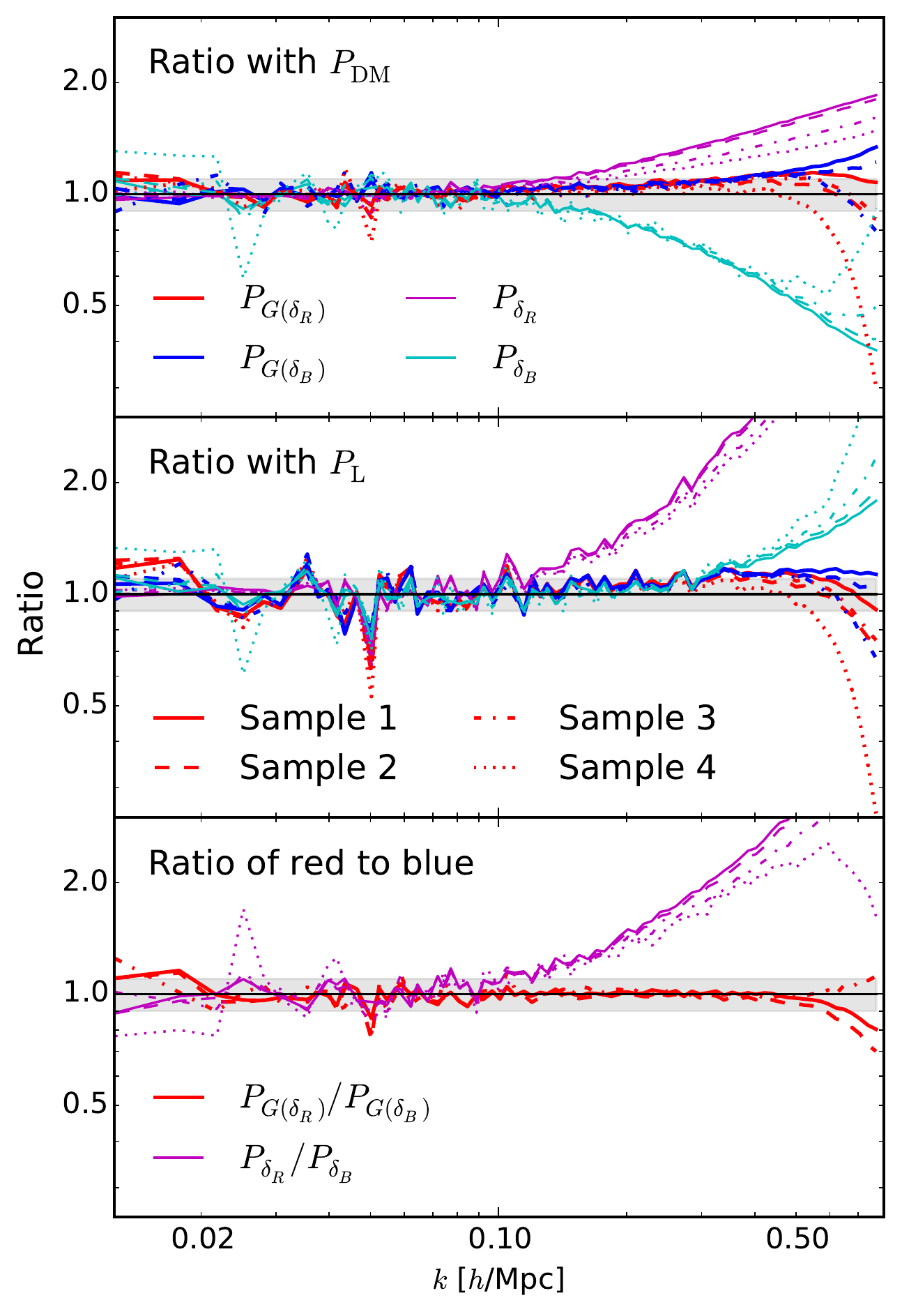}
}
\caption{Variation of results with respect to grid size (left panels) and sample definitions (right panels). Ratios of the galaxy power spectra of the usual density and transformed density fields to the underlying dark matter (top), linear power spectrum (middle), and the ratio between red and blue power spectra (bottom). In both panels: red and magenta lines show the Gaussianized and usual density power spectra of red galaxies. Blue and cyan lines show those of blue galaxies. Left panels: the different line styles show different grid sizes. Right panels: the different line styles show the different samples, summarized in Table \ref{tab:samples}. The transformed density power spectrum for the blue galaxies is not shown for Sample 4 (see text).}
\label{fig:pkratiogridsample}
\end{figure*}

The left panels of Fig. \ref{fig:pkratiogridsample} show real-space ratios of the galaxy power spectra of the usual density and transformed density fields to the underlying dark matter (top), initial power spectrum (middle), and the ratio between red and blue power spectra (bottom). Gaussianized power spectra for red and blue galaxies are shown with the red and blue lines, and the usual density power spectra are shown with magenta and cyan lines, respectively. The different line styles in each colour show 3 different cell sizes used in the CIC density assignment. The solid line shows a cell size of $2$ Mpc/$h$, which corresponds to a $256^3$ CIC grid in our simulation box. The dashed line is the $4$ Mpc/$h$ grid size that was used in the previous section, corresponding to a $128^3$ grid. The largest cell size shown here is $8$ Mpc/$h$, shown in the dot-dashed lines, which corresponds to a $64^3$ grid. 

In the usual density power spectra (magenta and cyan lines in Fig. \ref{fig:pkratiogridsample}), the cell size has almost no effect on these ratios, except near the Nyquist frequency of each grid, as expected. The shape of the Gaussianized power spectra, on the other hand, is in general dependent on the grid size chosen. For example, with very large cells, there are few empty cells, and the cells in the Gaussianized field populate the full Gaussian PDF. However, with very small cells, there are many empty cells, and after the transformation, the non-zero cells only populate a small fraction of the Gaussian PDF. Therefore, there is a tradeoff between the reducing the number of empty cells, which make the PDF of the Gaussianized field somewhat non-Gaussian, and reaching smaller scales with smaller cells.

For the ratio with dark matter (Fig. \ref{fig:pkratiogridsample}, top left panel), the grid size does affect the small-scale shape of the ratio between the Gaussianized spectra, but does not significantly change the scale to which there is agreement between the Gaussianized galaxy power spectra and the underlying dark matter spectrum. In all cases, the Gaussianized power spectra deviate from the dark matter power spectrum around $k\sim 0.3-0.4$ $h$/Mpc. 

For the ratio with the initial power spectrum (middle left panel), there is slightly more variation with cell size, with the $8$ Mpc/$h$ and $4$ Mpc/$h$ grids showing the best agreement with the initial power spectrum. For smaller cells, the Gaussianized power spectra begin to deviate at larger scales. This indicates that a larger grid size (more particles per cell) is preferable for this ratio.

For the ratio between the red and blue power spectra (bottom left panel), the cell size does not have a very large effect on the scale to which the power spectra agree. For the larger cells, the agreement extends to $k\sim 0.4-0.5$ $h$/Mpc, and for the larger cells it extends to the Nyquist frequency of the grid, $k=0.4$ $h$/Mpc.

The right panels of Fig. \ref{fig:pkratiogridsample} show how these ratios vary with sample selection. The right panels correspond to the same ratios as the left, with the same colour scheme, but the line styles now show the 4 samples given in Table \ref{tab:samples}. We use a $4$ Mpc/$h$ cell size for all power spectra in this figure. Note we have not included a measurement for the blue galaxies from Sample 4. As mentioned in Section \ref{sec:sec2sub2}, the shot noise correction scheme works well for a range of number densities and grid sizes. For Sample 4, the number density of the blue galaxies is ($\bar n=0.001$ (Mpc/$h$)$^{-3}$) and with a $128^3$ grid, the shot noise correction procedure is not accurate on the scales of interest.

For the usual density statistics in the right-hand panels of Fig. \ref{fig:pkratiogridsample} (cyan and magenta lines), the sample selection has an effect on the ratios shown. This is because the different stellar mass cuts give samples with different clustering properties, including nonlinear bias.

For the Gaussianized statistics, the sample selection has remarkably little effect on these ratios. The red galaxies in Sample 4 (red dotted line) show a deviation at larger scales than the other samples. We attribute this to the shot noise correction procedure, which is only accurate on scales below $k\sim 0.5$ $h$/Mpc for the number density of this sample ($\bar n=0.006$ (Mpc/$h$)$^{-3}$). Therefore, we do not expect agreement on scales smaller than $k\sim 0.5$ $h$/Mpc for this sample.

\begin{figure*}
\centering
\subfloat{
\includegraphics[width=0.485\textwidth]{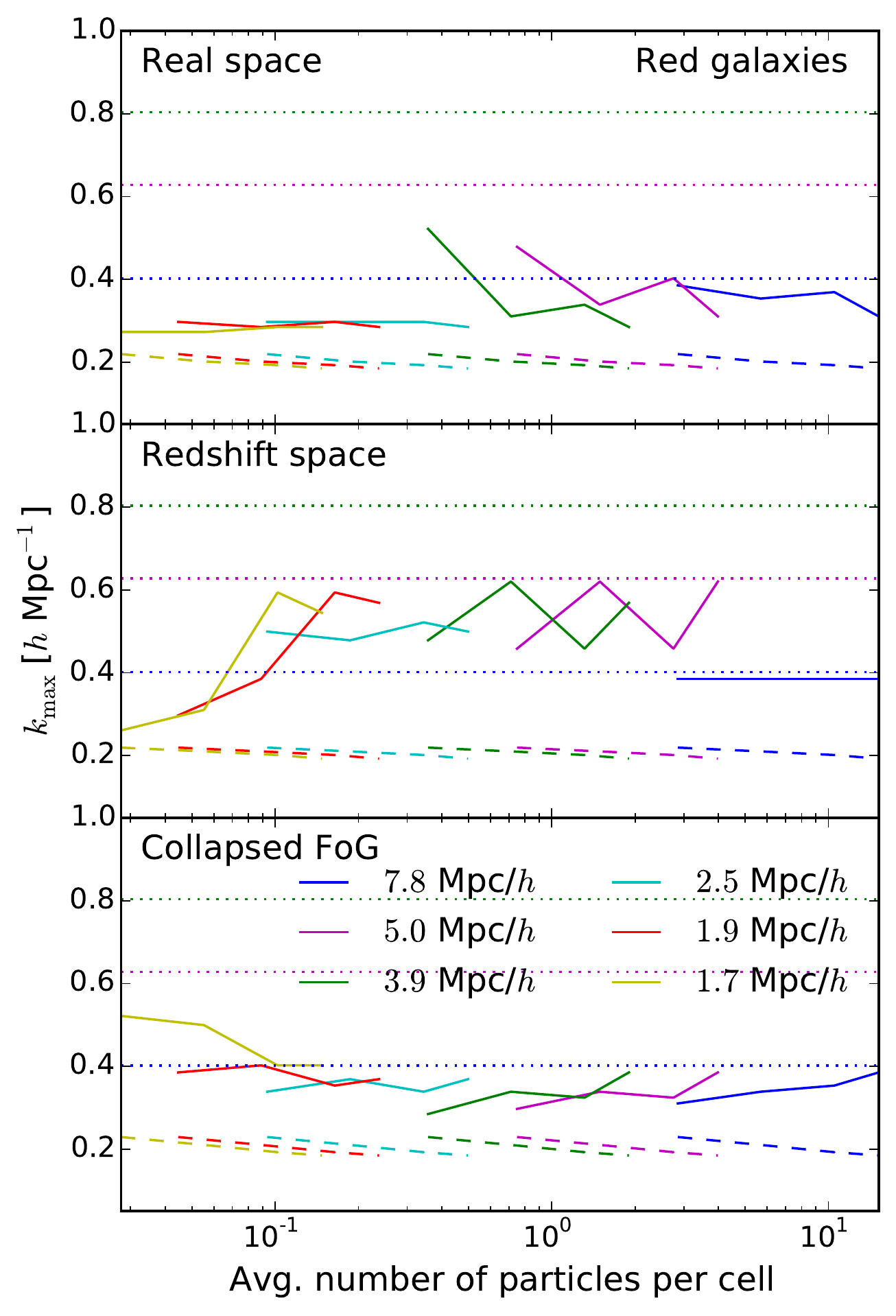}
}
\subfloat {
\includegraphics[width=0.485\textwidth]{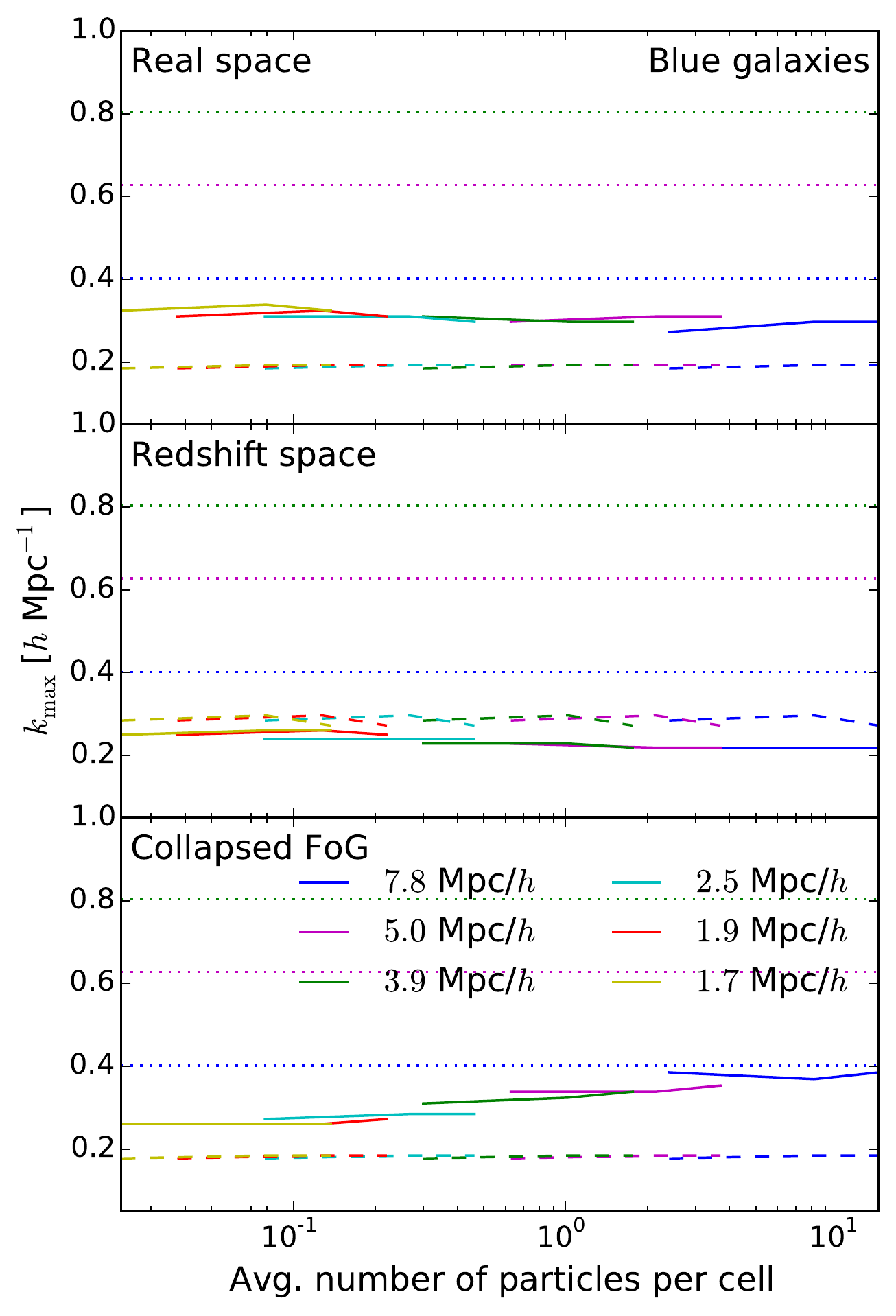}
}
\caption{Maximum wavenumber, $k_{\textrm{max}}$, for which the galaxy power spectrum $P_{\textrm{gal}}(k)$ provides a statistically adequate description of the underlying $P_{\textrm{DM}}(k)$, as qualified by Eq. \ref{eq:chisq}. Left (right) panel shows $k_{\textrm{max}}$ as a function of the average number of particles per cell for red (blue) galaxies and top to bottom shows real space, redshift space, and collapsed FoG space, respectively. Dashed lines show $k_{\textrm{max}}$ for the usual density power spectrum and solid lines show $k_{\textrm{max}}$ for the Gaussianized power spectrum. Nyquist frequencies for the largest 3 cell sizes are indicated as horizontal dotted lines.}
\label{fig:kmaxcell}
\end{figure*}

To quantify the agreement between the galaxy power spectra and the underlying dark matter (or initial) power spectrum, we use a reduced $\chi^2$ to test the goodness-of-fit of the dark matter spectrum to that of the galaxies as a function of $k$-bin:
\begin{align}
\chi^2_n\equiv\chi^2(k_n)&=\frac{1}{\nu} \sum_{i=0}^n \frac{\left(\ln P_{\textrm{gal}}(k_i) - \ln P_{\textrm{DM}}(k_i)\right)^2}{\sigma^2},
\end{align}
where $\nu$ is the number of degrees of freedom, and $\sigma^2$ is the variance of $\ln (P_{\textrm{G}}(k_i))$. For bin $n$ in the reduced $\chi^2$, the number of degrees of freedom is $n+1$. For simplicity, we assume that the underlying density field is Gaussian, implying that the variance in bin $i$ is $2/N_i$, where $N_i$ is the number of modes in the bin. While this assumption is certainly not true on small scales, we use this quantity only to estimate the scale of deviation of the power spectra, so it does not need to be extremely accurate.

With these assumptions, we arrive at the following expression for the reduced $\chi^2$:
\begin{align}
\chi^2_n&=\frac{1}{2(n+1)}\sum_{i=0}^n \left(\ln \frac{P_{\textrm{gal}}(k_i)}{P_{\textrm{DM}}(k_i)}\right)^2 N_i\label{eq:chisq}.
\end{align}
We define the quantity $k_{\textrm{max}}$ to be the highest $k$ where the reduced $\chi_n^2<1$. This corresponds to the scale at which the galaxy power spectrum $P_{\textrm{gal}}(k)$ no longer provides a statistically adequate description of the underlying $P_{\textrm{DM}}(k)$.

In order to test whether there is an optimal cell size for a given sample, we compute the $k_{\textrm{max}}$ values for the ratios of the usual and transformed power spectra with the underlying dark matter power spectrum in all 4 samples over a range of cell sizes. We check that $k_{\textrm{max}}$ is never larger than the shot noise correction scale for a given grid size and sample. Fig. \ref{fig:kmaxcell} shows $k_{\textrm{max}}$ for the usual density (dashed lines) and Gaussianized density (solid lines) cases, as a function of the average number of particles per cell for red galaxies (left panels) and blue galaxies (right panels). The different colours of solid and dashed lines represent different cell sizes used for each of the four samples given in Table \ref{tab:samples}. The horizontal dotted lines mark the Nyquist frequencies of the corresponding grids. The three vertical panels in Fig. \ref{fig:kmaxcell} show $k_{\textrm{max}}$ values in real space (top), redshift space (middle), and collapsed finger-of-god space (bottom).

These figures show that in the ratio of Gaussianized galaxy spectra to the Gaussianized dark matter spectrum (solid lines), there is not a strong dependence on number of particles per cell in most cases, but that a higher number of particles per cell is preferred in some cases. Apart from red galaxies in redshift space, there is no need to use cells smaller than $8$ Mpc/$h$, as $k_{\textrm{max}}$ never exceeds the Nyquist frequency of that grid. In the case of the red galaxies in redshift space, $k_{\textrm{max}}$ can be as high as $0.6$ $h$/Mpc if a smaller grid size is used, corresponding to an average of 1 particle per cell.

We also computed the $k_{\textrm{max}}$ values for the ratio with the initial power spectrum and for the ratio of red to blue power spectra in all cases. We found that for the ratio with the initial power spectrum, larger cells are optimal for both red and blue galaxies. In both cases, the agreement in real space extends to $k\sim 0.4$ $h$/Mpc for the $64^3$ grid. For the ratio between red and blue power spectra, the agreement can be extended to $k\sim0.6$ $h$/Mpc using a grid size corresponding to $\sim 1$ particle per cell in both real space and with collapsed fingers of god.

Overall, we find that for Gaussianized galaxy power spectra, a minimum cell size of about $4$ Mpc/$h$ is sufficient, as the agreement in shape between the galaxy spectra and dark matter never exceeds $k\sim 0.6$ $h$/Mpc. Using smaller cells tends to make the agreement worse. While the optimal cell size has a complicated dependence on number density and clustering, and may also depend on which ratio one is interested in, we find that a cell size that roughly corresponds to an average of 1 or more particles per cell is a good rule of thumb.

\section{Comparison to Clipping}
\label{sec:clipping}

In this section we compare the Gaussianization transform to the clipping transform \citep{simpsonClipping}, which has also been shown to reduce the effect of both nonlinearity and nonlinear bias in the galaxy power spectrum \citep{simpson2015}. First, we describe the clipping procedure in general and how we apply it to the red and blue galaxy samples in Sample 1 in both real and redshift space. We then discuss how the method and results compare to Gaussianization.

The basic idea of the clipping transform is to impose an upper limit on the density in the field under consideration. For a given density threshold, $\delta_0$, all cells with $\delta_i>\delta_0$ are set to the threshold value, $\delta_0$. Formally, the clipped density field ($\delta_c(\x)$) is related to the usual density field ($\delta(\x)$) through the following transformation:
\begin{align}
\delta_c(\x)&=\begin{cases}
\delta_0, & \delta(\x)>\delta_0\\
\delta(\x), & \delta(\x)\le \delta_0\notag.
\end{cases}
\end{align}
This suppresses the high-density peaks which contribute most to the nonlinear behaviour of the density field. As shown in \citet{simpsonClipping, simpson2015}, the statistics of the clipped density field can be modelled accurately to higher wavenumbers ($k\sim 0.5$ $h$/Mpc) than those of the usual density field. 

\citet{simpson2015} applied clipping to the full galaxy sample from the GAMA survey to constrain $f\sigma_8$. The various threshold values used were chosen such that the large-scale amplitude of the galaxy power spectrum was reduced by 30\% to 60\%, corresponding to removal of roughly 10\% to 20\% of the objects in the field. This balances between reducing the effects of nonlinearity and maintaining the signal to noise.

Here, we are interested in testing clipping on galaxy fields separated by colour. As the red and blue galaxy fields we are considering have very different large-scale linear biases ($b_{\textrm{blue}}=0.80$ and $b_{\textrm{red}}=1.55$ in Sample 1, as computed with Eq. \ref{eq:lsamp}), it is unclear what the optimal clipping threshold is, and whether it should be the same for both galaxy fields. Thus, we test a number of different threshold values in the two fields. See Table \ref{tab:clipping} for details of the various thresholds used, for Sample 1.

In order to model the resulting clipped galaxy power spectrum, we follow Equations 27-29 in \citet{simpson2015}, which relate the clipped galaxy power spectrum to the clipped dark matter power spectrum. Here, instead of using an analytic prediction for the nonlinear dark-matter power spectrum in real or redshift space, we use the known underlying dark matter field from the simulation. To model the power spectrum of a galaxy field clipped to a given large-scale amplitude, we scale the dark matter field by the measured linear bias of the galaxy field and apply clipping at a level that results in a clipped biased dark matter power spectrum with the same large-scale amplitude as the clipped galaxy power spectrum. In other words, our model for the clipped galaxy power spectrum is:
\begin{align}
\mathrm{Cl}[P_{\mathrm{gal}}(k)]=\mathrm{Cl}[b_{\mathrm{gal}} P_{\mathrm{DM}}(k)],\label{eq:clipmodel}
\end{align}
where the strength of the clipping operation, represented by $\mathrm{Cl}[\cdots]$, is defined by the reduction in large-scale amplitude of the resulting clipped power spectrum. In general, the density thresholds used (and percentage of mass removed) for the galaxy and linearly-biased dark matter fields may be different for each galaxy sample to achieve the same large-scale amplitude (see Table \ref{tab:clipping}). Note that the change is not necessarily huge in terms of mass fractions: a few percent, at most.

We estimate the clipped power spectrum as described in Section \ref{sec:sec2sub2} using an FFT of a $128^3$ CIC grid, after subtracting the mean of the clipped overdensity field to ensure that $\langle \delta_c \rangle =0$. As described in \citet{simpson2015}, the shot noise contribution to the clipped power spectrum can be modelled for clipped Gaussian fields as $P_s\simeq f_V^2/\bar n$, where $f_V$ is the fraction of the volume of the field lying below the clipping threshold. For the thresholds used here, $f_V$ is higher than about 85\% in all cases.

\begin{table*}
\centering
\caption{Summary of different clipping thresholds used for Sample 1 in real space. The first two columns give the large-scale amplitude of the clipped power spectra relative to the unclipped dark-matter power spectrum and unclipped blue power spectrum, respectively (see Eq. \ref{eq:lsamp} for definition of the large-scale amplitude). The middle four columns give the real-space density thresholds (and corresponding percentage of objects removed within parentheses) for the red galaxy field, the dark matter field scaled by the bias of the red galaxies, the blue galaxy field, and the dark matter field scaled by the bias of the blue galaxies. The final column gives notes about the chosen thresholds.}
\label{tab:clipping}
\begin{tabular}{c|c|c|c|c|c|l}
 $B_{\textrm{clip}}/B_{\textrm{DM}}$&$B_{\textrm{clip}}/B_{\textrm{blue}}$ &$\delta_0^{\textrm{red}}$  &$(b_{\textrm{red}}\delta^{\textrm{DM}})_0$ & $\delta_0^{\textrm{blue}}$ & $(b_{\textrm{blue}}\delta^{\textrm{DM}})_0$ & Notes \\
\hline \hline
1.0 &1.56& 9.8 & 5.4 & --- & --- & Red galaxies clipped to match the large-scale\\
 &&   (17.8\%) & (19.6\%) &   &  &amplitude of DM power spectrum \\
 \hline
 0.51& 0.8 & 3.95 & 2.05 & 3.85 & 10.5 & Red and blue galaxies clipped so large-scale\\
 && (31.9\%) & (35.8\%) & (3.1\%) & (2.2\%) &amplitudes are 80\% of unclipped blue power spectrum\\
  \hline
 0.26 & 0.4 & 1.7 & 0.7 & 1.32 & 1.6 &Red and blue galaxies clipped so large-scale\\
 && (44.8\%) & (51.1\%) & (15.8\%) & (10.7\%) &amplitudes are 40\% of unclipped blue power spectrum
\end{tabular}
\end{table*}

\begin{figure}
\centering
\includegraphics[width=0.485\textwidth]{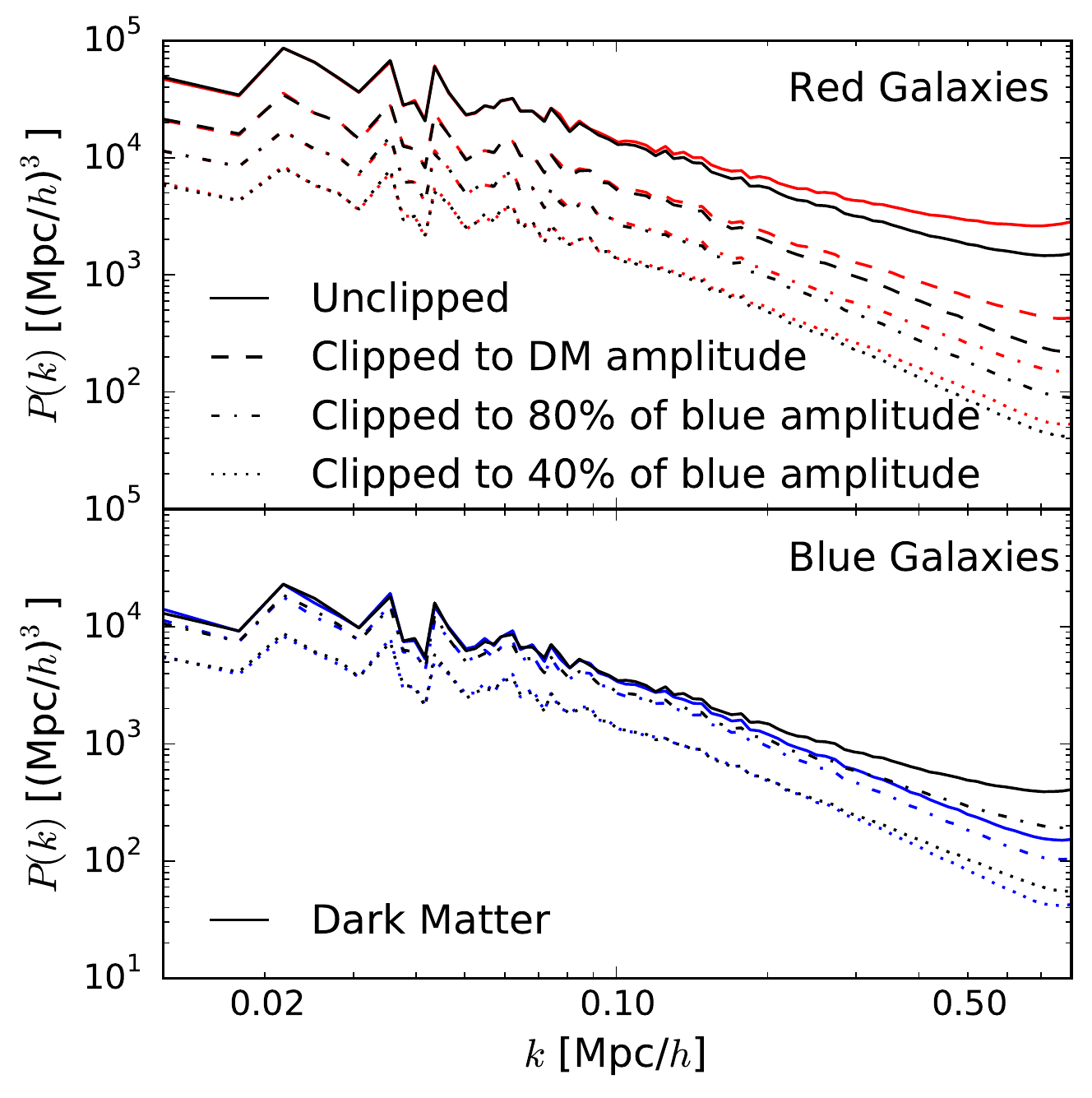}
\caption{The effect of clipping on red and blue galaxy power spectra in real space. Line styles indicate clipping level (see key), and top and bottom panels show red and blue galaxies, respectively. Black lines in both panels show power spectra of dark matter, linearly biased by the appropriate $b_{\mathrm{gal}}$, and clipped to the same large-scale amplitude as the corresponding galaxy power spectrum. Stronger clipping results in agreement on smaller scales between galaxy and dark matter power spectra.}
\label{fig:clip_pk}
\end{figure}

Fig. \ref{fig:clip_pk} shows the clipped and unclipped power spectra in real space. The top panel shows the unclipped (solid) and clipped (dashed, dot-dashed, and dotted) red galaxy power spectra (red lines) and corresponding clipped linearly-biased dark matter power spectra (black lines). The bottom panel shows the same for the blue galaxies. By construction, the large-scale amplitudes of the clipped galaxy and clipped (biased) dark matter power spectra are the same. It is clear from this figure that clipping reduces the small-scale power of both the galaxy and dark matter power spectra, bringing them closer to each other on small scales, as expected.

\begin{figure*}
\centering
\subfloat{
\includegraphics[width=0.485\textwidth]{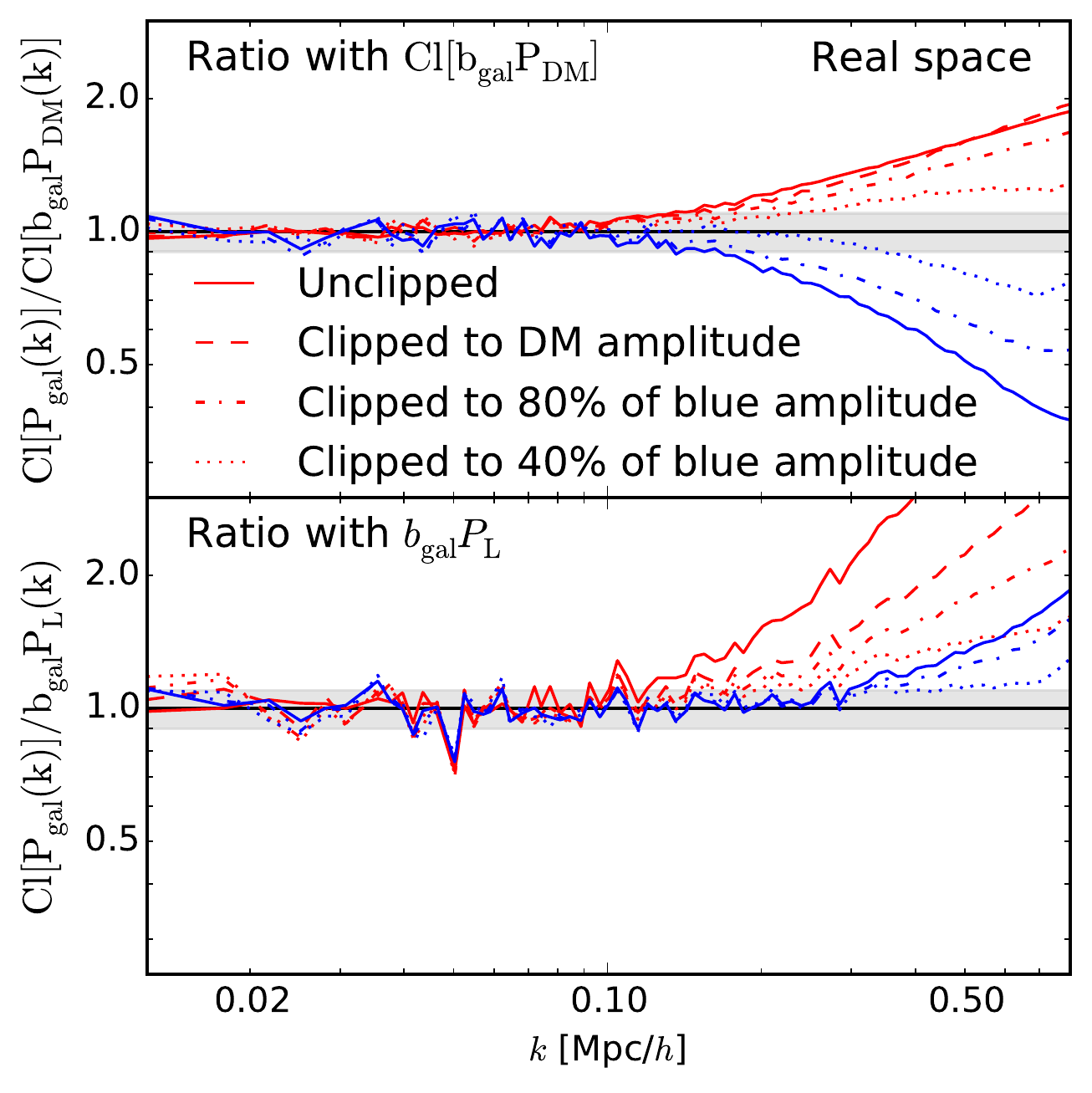}
}
\subfloat {
\includegraphics[width=0.485\textwidth]{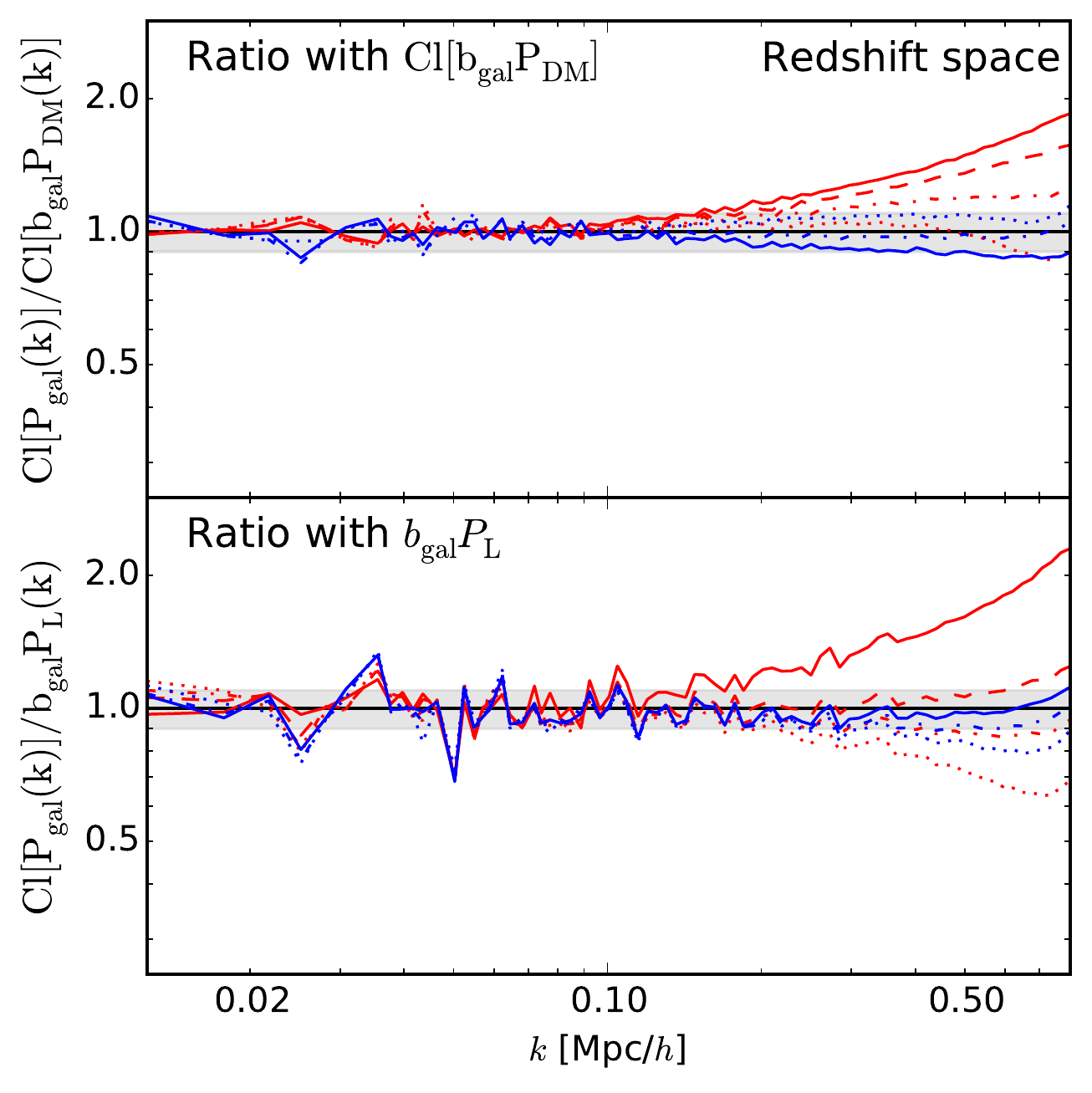}
}
\caption{Ratios between unclipped (solid) and clipped (dashed, dot-dashed, and dotted) galaxy power spectra and corresponding linearly-biased dark matter power spectra (top) or initial power spectrum (bottom). Left panels show the ratios in real space and right panels show ratios in redshift space. Shaded grey regions show the 10\% range around 1. See Eq. \ref{eq:clipmodel} and text for discussion of the clipping procedure and modelling. Stronger clipping reduces the effects of both nonlinearity and nonlinear galaxy bias. In redshift space the effect of clipping is more complicated than in real space.}
\label{fig:clip_pk_ratio}
\end{figure*}

This can be seen more clearly in Fig. \ref{fig:clip_pk_ratio}, which shows the ratios of the galaxy and dark matter power spectra in the top panels, and the ratios of the galaxy and initial power spectra in the bottom panels. The left panels of Fig. \ref{fig:clip_pk_ratio} show these ratios in real space and the right panels show the ratios in redshift space (monopole). Red and blue lines correspond to red and blue galaxy samples in all panels, and the various clipping thresholds are shown in different line styles. From this figure, we see that increasing the clipping strength improves the agreement between the clipped galaxy fields and both the (clipped biased) nonlinear dark matter power spectrum and the linear power spectrum. However, even at the strongest clipping level in the red galaxies, which removes nearly 45\% of the galaxies, the agreement with both the dark matter and linear power spectra only extends to $k\sim 0.2$ $h$/Mpc. The agreement is better in the case of blue galaxies at this clipping level, extending to $k\sim 0.3-0.4$ $h$/Mpc.

Note that when we repeat the clipping procedure in redshift space, we find our clipping thresholds relative to the redshift-space large-scale amplitudes. This means that for a given field, the density threshold used in real and redshift space may be different to achieve the same relative amplitude reduction. For example, to clip the red field to the dark-matter large-scale amplitude, we use a density threshold of $\delta_0^{\textrm{red}}=9.8$ in real space, and a threshold of $\delta_0^{\textrm{red}}=6.4$ in redshift space. As the right panels of Fig. \ref{fig:clip_pk_ratio} show, increasing the clipping strength improves the agreement on small scales between the galaxy power spectra and (clipped) dark matter, although the strongest clipping strength (dotted lines) does not seem to improve the agreement for blue galaxies. In the ratio with the initial power spectrum, the strongest clipping threshold overly suppresses both galaxy power spectra on small scales relative to the initial power spectrum.

From this analysis we can make several general comparisons with the Gaussianization transform, in both the applicability and effectiveness at removing nonlinearity and bias from the galaxy power spectra. There is a degree of ambiguity when applying clipping to galaxy fields that is not present in Gaussianization. For example, it is not clear what the optimal clipping threshold is for a given field, whether that threshold depends on the clustering of the field, and whether it is the same in real and redshift space. It is also not obvious whether the red and blue galaxy fields, clipped to a given large-scale amplitude, should agree with each other on small scales, whereas an explicit goal of the Gaussianization transform is to separate bias from intrinsic clustering resulting in fields with the same statistics. However, the shot noise contribution is much better behaved in the case of clipping, where it can even be reduced compared with the usual density field. With Gaussianization, the shot noise contribution is increased in a complicated way that depends on grid size and number density.

Overall, it is clear that there are subtleties and complications in both methods that must be well understood in order to achieve any benefits from their application. Both transforms appear to work reasonably well at removing nonlinearity and galaxy bias in real space. Clipping is more well behaved than Gaussianization in redshift space, provided the clipping strength is moderate.

\section{Conclusion}
\label{sec:conclusion}

The Gaussianization transform has previously been shown to be a promising method for restoring information to the 2-point statistics of the matter density field on quasi-linear scales. This is due to two effects: reduced covariance on small scales and the shape of the small-scale power spectrum. In this paper, we have studied the effectiveness of the transform to remove nonlinear galaxy bias and thus restore the small-scale shape of the underlying dark matter power spectrum when applied to differently biased galaxy samples from the semi-analytic model {\scshape galform}.

Overall, our results raise a few key points: the approximation that Gaussianization removes bias, which would imply that the Gaussianized statistics of differently biased fields agree, seems to hold in real space on quasi-linear scales ($k\sim 0.4-0.6$ $h$/Mpc). Gaussianization in real space also recovers the shape of the linear power spectrum on these scales ($k \sim 0.4$ $h$/Mpc).

Thus, the transform would seem to be an effective method for removing nonlinear bias and nonlinear gravitational evolution from the 2-point statistics of galaxies, and could simplify the modelling of modes up to $k\sim 0.4$ $h$/Mpc, likely improving constraints on cosmological parameters.

However, the galaxy-bias benefits observed in real space unfortunately do not extend to redshift space, for the most part. The Gaussianized statistics do not offer an improvement over the usual density statistics in estimating the linear-theory power spectrum in redshift space. While the shape of the Gaussianized blue galaxy power spectrum does not match that of the underlying dark matter in redshift space, we find that the shape of the Gaussianized red galaxy power spectrum in redshift space agrees very well ($k \sim 0.6$ $h$/Mpc) with that of the underlying dark matter. This suggests that Gaussianization of highly clustered galaxy samples, such as Luminous Red Galaxies (LRGs), may prove to be a useful method for recovering the nonlinear dark matter density field on quasi-linear scales. Further work is needed to understand why the Gaussianized statistics agree so well in this case. Another necessity is to investigate Gaussianization's effect on redshift-space clustering as a function of angle. It may be that Gaussianization is particularly adept at removing galaxy bias at angles where fingers of god do not dominate the signal, e.g. away from the line of sight.

We attribute the change in shape between the redshift-space Gaussianized matter and galaxy power spectra to the different behaviour of the galaxy samples on small scales. We test this by considering the redshift-space distribution with collapsed fingers of god. We find that in this case, the agreement is largely restored between the Gaussianized galaxy statistics and both Gaussianized dark matter and initial statistics. This provides a possible avenue for recovering the small-scale shape of the underlying dark matter and initial power spectra using Gaussianization.

We analyse the effect of sample selection and cell size on our results, and find that the agreement between the Gaussianized galaxy power spectra and underlying dark matter (and initial) power spectra in general depends on cell size and number density, as well as other factors, in a complicated way. However, in all cases using a cell size that corresponds to 1 or more galaxies per cell on average gives the best results and ensures that the shot noise is manageable.

We also compare the Gaussianization transform to the clipping transform, which has been shown to reduce the effects of nonlinearity and scale-dependent galaxy bias. We find that while clipping reduces the issue of shot noise and is less sensitive to resolution, there is ambiguity in the optimal clipping thresholds to use for each field. It is not clear how the optimal clipping threshold depends on the clustering of the field, and whether this threshold should be the same in real space and redshift space. Overall, we find that both clipping and Gaussianization can be effective at removing nonlinearity and nonlinear bias, at least in real space, if the subtleties of each procedure are carefully understood.

When applied appropriately, and if finger-of-god compression is available, the Gaussianization transform may be able to provide better fidelity to the dark matter and initial density power spectra, allowing for more modes to be included in analysis. Here we found that the agreement between the Gaussianized galaxy spectra and dark matter can extend to $k\sim 0.4-0.5$ $h$/Mpc, and similar agreement can be achieved between the Gaussianized galaxy spectra and the linear power spectrum at $z=0$. We expect the agreement to extend to even smaller scales for $z>0$, where nonlinearities are smaller.

Further benefits are likely to be seen in the reduced covariance, which we have not studied here. \citet{neyrinck2011b} used galaxy samples somewhat different from those used here, particularly not separated by colour, from the Millennium simulation to show that Gaussianization generally greatly reduces error bars in galaxy power spectra on small scales. The smaller error bars it provides can lead to smaller errors on any cosmological or galaxy-formation parameter to which galaxy power spectra may be sensitive, such as the power-spectrum tilt and neutrino masses.

\section*{Acknowledgements}
We thank Fergus Simpson for useful discussions. This work was supported by the Science and Technology Facilities Council (ST/L00075X/1). MN is grateful for financial support from a grant in Data-Intensive Science from the Gordon and Betty Moore and Alfred P. Sloan Foundations. PN acknowledges the support of the Royal Society through the award of a University Research Fellowship, the European Research Council, through receipt of a Starting Grant (DEGAS-259586) and support of the Science and Technology Facilities Council (ST/L00075X/1). 

This work used the DiRAC Data Centric system at Durham University, operated by the Institute for Computational Cosmology on behalf of the STFC DiRAC HPC Facility (www.dirac.ac.uk). This equipment was funded by BIS National E-infrastructure capital grant ST/K00042X/1, STFC capital grant ST/H008519/1, and STFC DiRAC Operations grant ST/K003267/1 and Durham University. DiRAC is part of the National E-Infrastructure.

\bibliographystyle{mnras}
\bibliography{gaussbib}

\end{document}